%% file: main.tex
\documentclass[11pt, a4paper, logo, copyright, nonumbering]{fdtn_template/fdtn}

\input{packages}
\input{commands}

\makeatletter
\def\@BTrule[#1]{%
  \ifx\longtable\undefined
    \let\@BTswitch\@BTnormal
  \else\ifx\hline\LT@hline
    \nobreak
    \let\@BTswitch\@BLTrule
  \else
     \let\@BTswitch\@BTnormal
  \fi\fi
  \global\@thisrulewidth=#1\relax
  \ifnum\@thisruleclass=\tw@\vskip\@aboverulesep\else
  \ifnum\@lastruleclass=\z@\vskip\@aboverulesep\else
  \ifnum\@lastruleclass=\@ne\vskip\doublerulesep\fi\fi\fi
  \@BTswitch}
\makeatother

\addto\extrasenglish{
}

 {\begin{list}{}%
         {\setlength{\leftmargin}{#1}}%
         \item[]%
 }
 {\end{list}}

\reportnumber{001} 

\input{fdtn_template/commands}

\definecolor{customlink}{HTML}{A9336A}
\renewcommand{\today}{}

\title{
    \vspace{-1em}
    {\textcolor{antares!300}{Antares:}}
    {Foundation Models for Agentic Vulnerability Localization}
}

\author{
    Supriti Vijay$^{*,1}$, 
    Aman Priyanshu$^{*,1}$, 
    Didier Chapoteau$^{1}$, 
    Arthur Goldblatt$^{1}$, 
    Jianliang He$^{1, 2, \dagger}$, 
    Kimia Majd$^{1}$, 
    Fraser Burch$^{1}$, 
    Baturay Saglam$^{1, 2, \dagger}$, 
    Takahiro Matsumoto$^{1}$, 
    Zhuoran Yang$^{1, 2, \dagger}$, 
    Amin Karbasi$^{1}$ \\
\vskip 0.1in
\begin{center}
\small
\opendata~\href{https://huggingface.co/collections/fdtn-ai/antares}{Models}\quad
\opencode~\href{https://github.com/cisco-foundation-ai/antares-cli}{CLI}\quad
{\textcolor{red!50!black}{\faGlobe}}~\href{https://cisco-foundation-ai.github.io/antares/}{Website}
\end{center}
\vskip -0.15in
}

\renewcommand{\firstpagefootnote}{%
  {\fontsize{8pt}{9pt}\selectfont
  $^{1}$Foundation AI--Cisco Systems Inc. \quad
  $^{2}$Yale University \quad
  $^{\dagger}$Work done while at Foundation AI \\ $^{*}$Equal Contribution. Corresponding authors: \texttt{\{suprivij,ampriyan\}@cisco.com}}%
}

\begin{document}

\input{content/main}

\end{document}

%% file: packages.tex
\usepackage[numbers, sort]{natbib}  
\usepackage{dblfloatfix}
\usepackage{ulem}
\usepackage{caption}
\usepackage{dramatist}
\usepackage{xspace}
\usepackage[table]{xcolor}
\definecolor{antares}{HTML}{ffd3b2}   
\definecolor{antares-sft}{HTML}{FFF0D9} 
\definecolor{antares-3b}{HTML}{f28c17}
\definecolor{gpt}{HTML}{C95689} 

\definecolor{custommustard}{HTML}{ffecec}
\definecolor{customred}{HTML}{ffcccc}
\definecolor{darkred}{HTML}{ff9797}

\definecolor{darkbrown}{HTML}{a25c22} 
\definecolor{mushyorange}{HTML}{efa802}
\definecolor{peach}{HTML}{fae0c7} 
\definecolor{mushygreen}{HTML}{bdc363}
\definecolor{magenta}{HTML}{dc426d}

\usepackage{tcolorbox}
\tcbuselibrary{most}
\usepackage{listings}
\usepackage{xcolor}
\usepackage{hyperref}

\lstdefinestyle{appendixlisting}{
    basicstyle=\ttfamily\small,
    breaklines=true,
    breakatwhitespace=false,
    columns=fullflexible,
    keepspaces=true,
    showstringspaces=false,
    numbers=none,
    frame=none
}

\newtcblisting{promptbox}[1][]{
    enhanced,
    breakable,
    listing only,
    listing options={style=appendixlisting},
    colback=antares-sft!10,
    colframe=antares-3b,
    boxrule=0.6pt,
    arc=1mm,
    left=1.5mm,
    right=1.5mm,
    top=1mm,
    bottom=1mm,
    before skip=8pt,
    after skip=8pt,
    fonttitle=\bfseries,
    title={Evaluation System Prompt},
    #1
}

\newtcblisting{toolbox}[1][]{
    enhanced,
    breakable,
    listing only,
    listing options={style=appendixlisting},
    colback=antares-sft!10,
    colframe=antares-3b,
    boxrule=0.6pt,
    arc=1mm,
    left=1.5mm,
    right=1.5mm,
    top=1mm,
    bottom=1mm,
    before skip=8pt,
    after skip=8pt,
    fonttitle=\bfseries,
    title={Tool Definition},
    #1
}

\newtcolorbox{observationbox}[1][]{
    enhanced,
    breakable,
    colback=orange!5,
    colframe=orange!65!black,
    boxrule=0.7pt,
    arc=1mm,
    left=2mm,
    right=2mm,
    top=1.5mm,
    bottom=1.5mm,
    before skip=8pt,
    after skip=8pt,
    fonttitle=\bfseries,
    title={Observation},
    #1
}

\usepackage{pgf-pie}
\usepackage{pgfplots}
\usepackage{tikz}
\usepackage{xcolor}

\pgfplotsset{compat=1.18}
\usepackage{xcolor}

\definecolor{fa1}{HTML}{FFBC8D}
\definecolor{fa2}{HTML}{FFC28C}
\definecolor{fa3}{HTML}{FFD08E}
\definecolor{fa4}{HTML}{FFB2A1}
\definecolor{fa5}{HTML}{F59CC4}
\definecolor{fa6}{HTML}{E7A6E5}
\definecolor{fa7}{HTML}{D8B3F1}
\definecolor{fa8}{HTML}{CCB5F5}
\definecolor{fa9}{HTML}{E8D5FB}

\newcommand{\slice}[6]{%
  \filldraw[
    fill=#6,
    draw=white,
    line width=1.2pt
  ]
  (0,0) -- (#1:2.8)
  arc[start angle=#1,end angle=#2,radius=2.8]
  -- cycle;

  \node[font=\small] at (#3:#4) {#5};
}

\usepackage{multirow}
\usepackage{xltabular}
\usepackage{longtable}
\usepackage{hyperref}
\usepackage{amsfonts}
\usepackage{amsmath}
\usepackage{amssymb}
\usepackage{lineno}
\usepackage{adjustbox}

\usepackage[bottom]{footmisc}

\usepackage{fontawesome5}
\usepackage{simpleicons}

\usepackage{setspace}

\usepackage{graphicx}
\usepackage{array, booktabs, tabularx}
\hypersetup{hidelinks=true, colorlinks=true, citecolor=blue, linkcolor=black, urlcolor=black}
\usepackage[textsize=scriptsize]{todonotes}
\usepackage{algorithm}
\usepackage{algpseudocode}
\usepackage{listings}
\usepackage{xcolor}
\usepackage{needspace}
\Needspace{8\baselineskip}

\usepackage{makecell}
\usepackage{inconsolata}

\usepackage{fvextra}
\usepackage{pgfplots}
\pgfplotsset{compat=1.18}
\usetikzlibrary{positioning, shapes.geometric, fit, backgrounds} 
\usepackage{pgf-pie}
\usepackage{fancyvrb}
\usepackage{mdframed}

\usepackage{arydshln}
\usepackage{threeparttable}
\usepackage{subcaption}
\usepackage{wrapfig}

\usepackage{dsfont}
\usepackage{array}
\usepackage{tabularx}
\usepackage{xcolor}

\usepackage{lipsum}
\usepackage{multicol}

\usepackage[utf8]{inputenc} 
\usepackage[T1]{fontenc}    
\usepackage{hyperref}       
\usepackage{url}            
\usepackage{booktabs}       
\usepackage{amsfonts}       
\usepackage{nicefrac}       
\usepackage{microtype}      
\usepackage{xcolor}         

\usepackage{environ}      
\tcbuselibrary{listings,breakable}

%% file: commands.tex
\definecolor{supriti-color}{HTML}{7881F2}

\definecolor{arthur-color}{HTML}{9B1C31}

\newcommand{\vlb}{VLoc Bench\xspace}

\newcommand{\opendata}[1][1em]{%
  \raisebox{-0.2\height}{\includegraphics[height=#1]{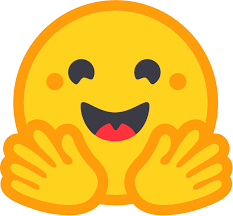}}%
}
\newcommand{\opencode}[1][1em]{%
  \raisebox{-0.2\height}{\includegraphics[height=#1]{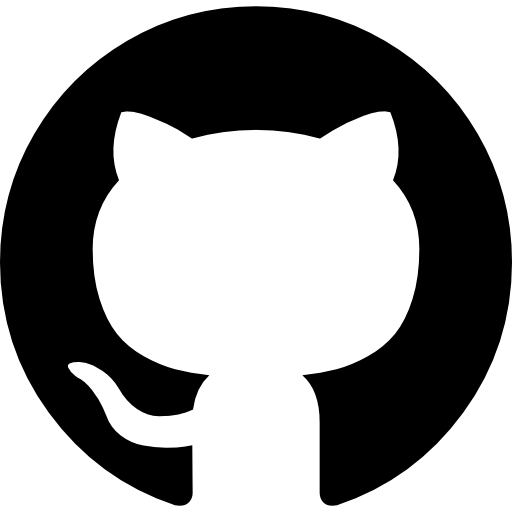}}%
}

%% file: fdtn_template/commands.tex
\renewcommand{\phi}{\varphi}

\renewcommand{\geq}{\geqslant}

\renewcommand{\epsilon}{\varepsilon}
\renewcommand{\imath}{\mathrm{i}}

\newlength{\restsubwidth}
\newlength{\restsubheight}
\newlength{\restsubmoreheight}
\newcommand{\rest}[2]{%
        \settowidth{\restsubwidth}{\ensuremath{#2}}
        \settoheight{\restsubheight}{\ensuremath{{}_{#2}}}
        \ensuremath{{#1\hskip 0.5pt}_{\vrule\kern2pt\parbox[b][%
        4pt][b]{\the\restsubwidth}{%
                        \ensuremath{{}_{#2}}}}}
        }

%% file: content/main.tex
\input{content/abstract}
\maketitle

\begin{figure}[h!]
    \centering
    \includegraphics[width=0.8\linewidth]{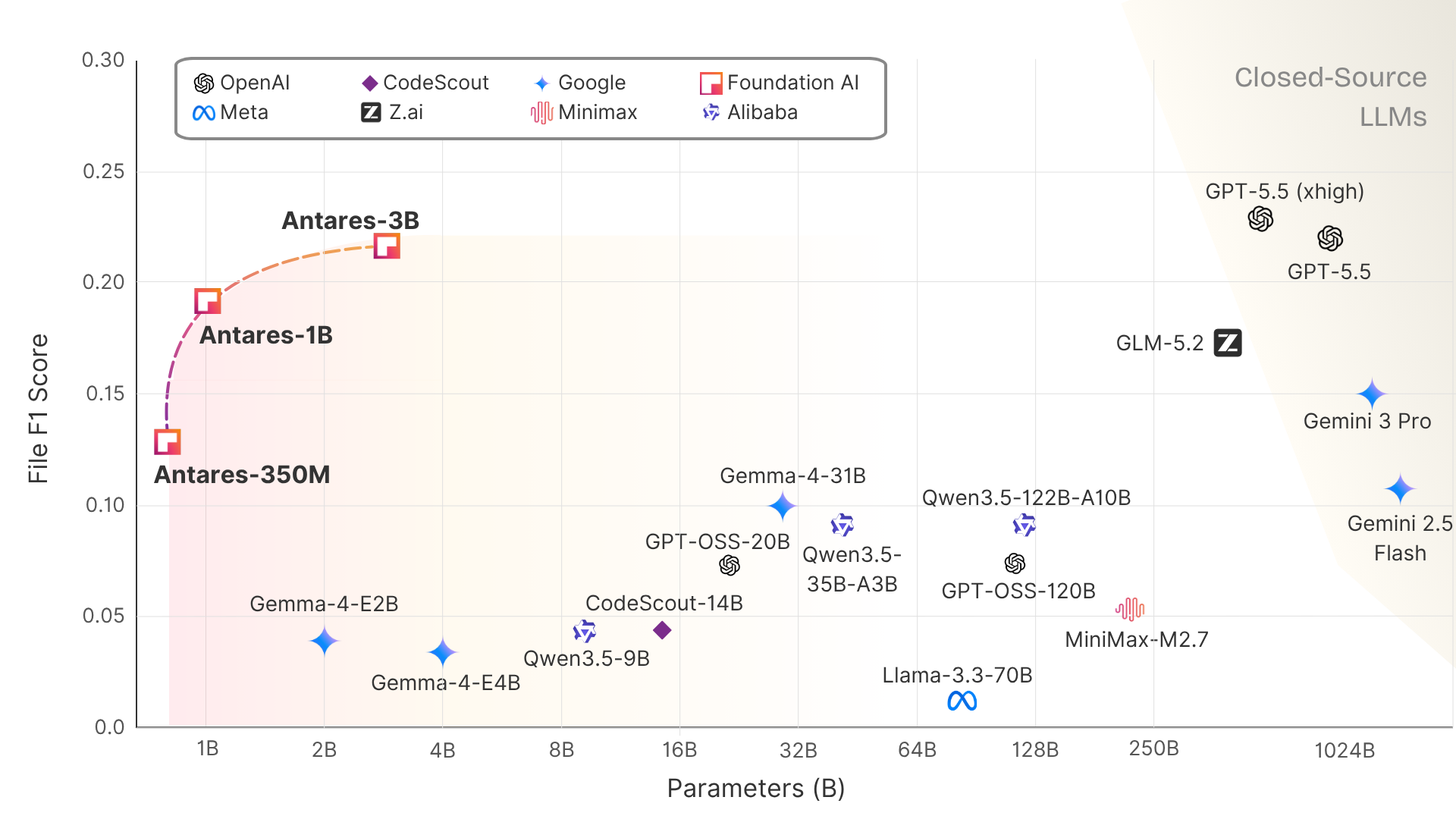}
    \caption{F1 score versus model size on Vulnerability Localization Benchmark (\vlb{}) \cite{manuscript-vlb}, a repository-scale benchmark comprising 500 tasks across 290 unique real-world repositories, where models receive only a CWE category description and must identify vulnerable implementation files in real codebases. Antares models form the Pareto frontier among evaluated models, achieving the strongest localization quality at small parameter scales. Antares-3B reaches near-frontier closed-source performance while remaining orders of magnitude smaller than GPT-5.5 and Gemini-family baselines.}
    \label{fig:main_figure}
\end{figure}

\clearpage
\setcounter{tocdepth}{2}
\tableofcontents
\clearpage

\input{content/1-introduction}

\input{content/2-related-work}
\input{content/3-overview}

\input{content/4-data}
\input{content/5-training-pipeline}
\input{content/6-inference-pipeline}
\input{content/7-evaluation}
\input{content/8-results}
\input{content/9-error_analysis}
\input{content/10-safety}
\input{content/11-conclusion}
\input{content/acknowledgements}

\bibliographystyle{plainnat}
\bibliography{bibliography}

\appendix
\input{content/appendix}

%% file: content/abstract.tex
\begin{abstract}
Vulnerability localization is a fundamental step in software security, requiring models to reason over large codebases and iteratively identify vulnerable implementations. We present {\bf Antares}, a family of compact language models (350M, 1B, and 3B parameters) for agentic vulnerability localization. Based on IBM Granite base models, Antares is trained through a two-stage pipeline that combines supervised fine-tuning on cybersecurity reasoning and repository exploration data with reinforcement learning from verifiable rewards over vulnerable repositories. Across extensive evaluations, Antares-3B approaches GPT-5.5 while outperforming open-weight models over 200$\times$ larger in size. The Antares family further enables fast, low-cost local inference, completing a full 500-task evaluation sweep in approximately 15 minutes on a single H100 GPU, corresponding to an amortized evaluation time of under 2 seconds and less than \$0.002 per task. 

\end{abstract}

%% file: content/1-introduction.tex
\section{Introduction}

Modern software repositories are too large, modular, and dependency-rich for vulnerability remediation to begin with manual inspection alone \cite{Sabetta_2018}. Once a vulnerability is disclosed, the first operational question is not whether the weakness exists in the abstract, but where the vulnerable implementation lives. Accurately localizing that code is the step that enables patching, triage, regression testing, and downstream security review \cite{Sabetta_2018}. Yet repository-scale vulnerability localization remains difficult because the relevant evidence is rarely contained in a single function or file. It is distributed across imports, call paths, framework conventions, configuration boundaries, and implementation-specific idioms \cite{wang2024reposvulrepositorylevelhighqualityvulnerability, guo2025repoauditautonomousllmagentrepositorylevel}.

Human security researchers solve this problem interactively. They do not read an entire repository from top to bottom. They form hypotheses from the vulnerability class, search for likely entry points, inspect candidate files, follow call chains, compare naming conventions, and revise their search as new evidence appears. Vulnerability localization is therefore not simply a static code understanding task; it is an agentic reasoning problem over a live software environment.

Existing approaches only partially address this setting. Static analysis tools such as CodeQL \cite{github2025codeql}, SonarQube \cite{sonarqube2025}, and Semgrep \cite{semgrep2025} provide scalable rule-based detection, but their effectiveness is limited by predefined patterns and the coverage of their analysis front ends. Recent project-level and agentic vulnerability detection systems move beyond isolated function classification but often fall into one of two regimes: either they rely on static-analysis front ends to surface candidate locations, inheriting the recall limits of those tools, or they wrap frozen frontier models in search scaffolds, incurring high cost without owning the underlying localization policy \cite{nie2025vulnllmrspecializedreasoningllm, wang2025vulagenthypothesisvalidationbasedmultiagent,tsigkourakos2026qrsrulesynthesizingneurosymbolictriad,charoenwet2026agenticscrautonomousagenticsecure,xi2026tracelinellmagent,liu2026synthesizingmultiagentharnessesvulnerability}. In both cases, the system does not learn end-to-end how to search a repository from a bare vulnerability description.

We introduce \textbf{Antares}, a family of compact language models trained specifically for agentic vulnerability localization. Given only a CWE category description and read-only terminal access to a repository, Antares autonomously searches the codebase, inspects files, gathers evidence, and submits the vulnerable implementation paths. Unlike systems that depend on pre-extracted context, SAST-generated candidates, crash traces, or external frontier APIs, Antares performs localization end-to-end from the repository itself.

Antares consists of 350M, 1B, and 3B parameter models initialized from IBM Granite checkpoints and post-trained through a two-stage pipeline. Supervised fine-tuning teaches cybersecurity reasoning, repository exploration, and terminal interaction. Reinforcement learning then optimizes complete multi-turn trajectories using verifiable file-level localization rewards. This training setup directly rewards the behavior required at deployment time: searching strategically, verifying candidates, and submitting vulnerable files under a fixed terminal budget.

We evaluate Antares on Vulnerability Localization Benchmark (\vlb{}) \cite{manuscript-vlb}, a repository-scale benchmark comprising 500 tasks drawn from 290 unique real-world vulnerable repositories. All models are evaluated under the same constrained agent protocol: read-only Docker sandbox, no network access, a fixed terminal-command budget, and only the CWE category description as input. This setting tests whether a model can act like a security localization agent rather than merely classify a preselected code snippet. To test whether this policy transfers beyond security, we additionally evaluate general issue-driven code localization in \autoref{app:additional_benchmarks}. There, Antares-3B remains competitive with dedicated CodeScout models trained specifically for SWE-Bench localization, approaching the 4B CodeScout baseline despite being trained for vulnerability localization rather than issue-resolution file localization.

Our results, summarized in Figure~\ref{fig:main_figure}, show that targeted post-training can matter more than raw model scale. Antares-3B reaches 0.223 File F1, approaching GPT-5.5 while outperforming substantially larger open-weight models, including GLM-5.2. Antares-1B achieves the highest recall among all evaluated systems, and even Antares-350M outperforms several larger general-purpose baselines. Behavioral analysis further shows that reinforcement learning induces a search--verify--refine strategy rather than generic repository browsing, while reducing run-to-run variance across model scales.

Finally, Antares is designed for deployment constraints that matter in security workflows. We expose Antares through a local CLI for file-level vulnerability localization. The deployment preserves the benchmark's CWE-conditioned task, default inspection budget, and ranked file-submission protocol while adding repository isolation and machine-readable reporting. When inference is hosted within the user's trust boundary, proprietary source code need not be sent to third-party APIs. In our evaluation setup, the full 500-task \vlb{} evaluation completed in approximately 15 minutes on a single H100 GPU, corresponding to under two seconds and less than \$0.002 per task. The CLI provides human-readable and structured outputs, including SARIF for code-scanning integration, and supports CI/CD, security triage, and closed-network or air-gapped deployment when the model weights and inference endpoint are hosted locally.

%% file: content/2-related-work.tex
\section{Related Work}

\subsection{Static Analysis and ML-Based Vulnerability Detection}

Traditional vulnerability detection relies on static analysis tools such as CodeQL \cite{github2025codeql}, SonarQube \cite{sonarqube2025}, and Semgrep \cite{semgrep2025} that identify security flaws through predefined patterns and dataflow rules. While these tools have become standard in development workflows, comprehensive evaluations consistently expose fundamental limitations including susceptibility to evasion techniques, inability to reason about novel vulnerability classes, and poor adaptation to unfamiliar codebases \cite{ami2024falsenegative}. Machine learning approaches, from graph neural networks to large language models fine-tuned for vulnerability detection, improve recall on benchmark datasets but exhibit significant precision instability across projects and remain constrained to single-pass analysis over fixed code snippets \cite{tihanyi2025vulndetection}. Recent project-level and agentic vulnerability detection systems move beyond isolated functions, but often rely on static-analysis front ends, pre-extracted contexts, crash traces, or frozen frontier models rather than end-to-end learned localization from a bare repository \cite{nie2025vulnllmrspecializedreasoningllm, wang2025vulagenthypothesisvalidationbasedmultiagent,tsigkourakos2026qrsrulesynthesizingneurosymbolictriad,charoenwet2026agenticscrautonomousagenticsecure,xi2026tracelinellmagent,liu2026synthesizingmultiagentharnessesvulnerability}. These limitations are particularly acute in repository-scale settings where vulnerability context spans multiple files and successful localization requires iterative navigation rather than one-shot classification.

\subsection{Security-Specialized Language Models}

Rather than improving detection tools directly, an alternative line of work trains language models to internalize security knowledge. Lily-Cybersecurity-7B \cite{jiang2023lily}, DeepHat-V1-7B \cite{deephat2025}, Foundation-Sec-8B \cite{kassianik2025foundationsec-base,weerawardhena2025foundationsec-instruct}, and Primus \cite{yu2025primus} demonstrate that continued pretraining or fine-tuning on security corpora covering vulnerability assessment, threat intelligence, and penetration testing can effectively transfer domain knowledge to models ranging from 7B to 13B parameters. Foundation-Sec-8B-Reasoning \cite{yang2026foundationsec-reasoning} extends this line through GRPO with verifiable rewards, producing the first open-source reasoning model for cybersecurity. These models excel at classification and structured reasoning but cannot act on their knowledge within real software environments, lacking the ability to navigate repositories, execute terminal commands, or iteratively explore codebases to locate vulnerable implementations.

\subsection{Agentic Information Retrieval and Code Localization}

Classical dense retrieval systems \cite{karpukhin2020dpr,wang2024e5} optimize a single query and assume the right evidence is surfaceable in one retrieval pass. Reasoning-aware retrievers \cite{shao2025reasonir,das2025rader} and reinforced query rewriting systems \cite{qin2025tongsearch} improve intent alignment by conditioning on chain-of-thought traces or learning reformulations against retrieval feedback, but remain single-turn in their search strategy. Interactive and agentic retrieval methods \cite{jin2025searchr1,jiang2025deepretrieval,zheng2025deepresearcher,li2025webthinker} address this by training LLMs to issue tool calls, read results, and iteratively refine queries. Think Before You Retrieve \cite{vijay2025thinkbeforeretrieve} further demonstrated that compact models (350M to 1.2B) can learn dynamic multi-turn retrieval strategies through turn-level GRPO rewards, outperforming larger specialized systems despite being 200 to 400 times smaller.

The proliferation of LLM-powered coding agents \cite{anthropic2026claudecode,openai2026codex,cursor2026cursor} has driven demand for dedicated code localization and context-building capabilities within these systems. SWE-grep \cite{pan2025swegrep} and Composer \cite{cursor2025composer} demonstrate that reinforcement learning produces emergent efficiency behaviors including parallel tool calls and search-heavy exploration strategies that match frontier retrieval accuracy at substantially higher throughput. CodeScout \cite{sutawika2026codescout} formalizes this direction by training code search agents purely with GRPO on SWE-Bench \cite{jimenez2024swebench} using file-level F1 as the reward signal across scales from 1.7B to 14B parameters, while FastContext \cite{zhang2026fastcontext} trains exploration subagents that return compact file-and-line citations to a main solving agent. Rather than targeting general software engineering, we extend this paradigm to terminal-based repository exploration specialized for vulnerability localization.

\subsection{Reinforcement Learning for Agentic Models}

Several concurrent works advance reinforcement learning for agentic language models at frontier scale. DeepSeek-R1 \cite{guo2025deepseekr1} pioneered GRPO \cite{shao2024deepseekmath} for reasoning, demonstrating that group-relative advantages with binary rewards can train explicit reasoning traces without learned reward models. The Qwen3 series \cite{yang2025qwen3} established a multi-stage recipe of SFT cold start followed by GRPO and strong-to-weak distillation. Large-scale agentic RL has since been applied in Kimi K2 \cite{kimik2026}, GLM-4.5 \cite{glm45team2025}, and MiniMax-M2 \cite{minimax2026m2}, each training models with hundreds of billions of parameters on agentic trajectories from real and synthetic environments. We adopt the same GRPO formulation and two-stage approach but apply it to multi-turn agent trajectories in the security domain at compact model scales rather than frontier-scale general reasoning.

\subsection{Agentic Cybersecurity Benchmarks}

Existing agentic benchmarks evaluate either general software engineering or offensive security capabilities but not repository-scale vulnerability localization. SWE-Bench \cite{jimenez2024swebench}, SWE-Bench Verified \cite{openai2024swebenchverified}, and SWE-Bench Pro \cite{deng2025swebenchpro} test issue resolution across Python repositories without requiring security-specific reasoning. $\tau$-bench \cite{yao2024taubench} and $\tau^2$-bench \cite{barres2025tau2bench} extend agentic evaluation to tool-agent-user interaction and dual-control environments, respectively, testing policy adherence without security-specific reasoning. NYU CTF Bench \cite{shao2024nyuctf}, CyberGym \cite{wang2026cybergym}, and ExploitGym \cite{wang2026exploitgym} evaluate offensive capabilities assuming the vulnerability location is already known, while CTIBench \cite{alam2024ctibench} tests security knowledge through classification tasks without any agentic interaction. This gap motivates \vlb{} \cite{manuscript-vlb}, a 500-task benchmark requiring models to simultaneously navigate unfamiliar codebases efficiently and recognize vulnerability patterns associated with specific CWE categories, a combination none of the above benchmarks evaluate.

%% file: content/3-overview.tex
\section{The Antares Family}

Antares is a family of compact language models designed for repository-scale vulnerability localization. Unlike conventional code generation models, Antares is trained to operate as an interactive software security agent capable of reasoning over complete repositories. Given a CWE-ID and its generic category description alongside read-only access to a software repository through a constrained terminal interface, the model autonomously explores the repository, gathers evidence, and identifies the source files containing the vulnerability.

The Antares agentic system consists of two tightly integrated components: a cybersecurity-specialized language model and an execution environment that exposes a restricted set of terminal operations. During inference, the model interacts with the repository through standard command-line utilities, including repository search, file inspection, and directory navigation. This interaction enables Antares to incrementally construct an understanding of unfamiliar codebases instead of relying solely on a fixed context window. To support deployment across diverse computational environments, Antares is developed at three model scales (350M, 1B, and 3B parameters) that share a common architecture and post-training pipeline.

Rather than specializing solely for code generation or cybersecurity question answering, Antares is optimized for the complete vulnerability localization workflow. This includes interpreting CWE category descriptions, planning repository exploration strategies, identifying relevant implementation components, synthesizing evidence across multiple files, and producing the final localization prediction. Throughout this report, we evaluate how these capabilities emerge through supervised fine-tuning and reinforcement learning over interactive terminal trajectories.

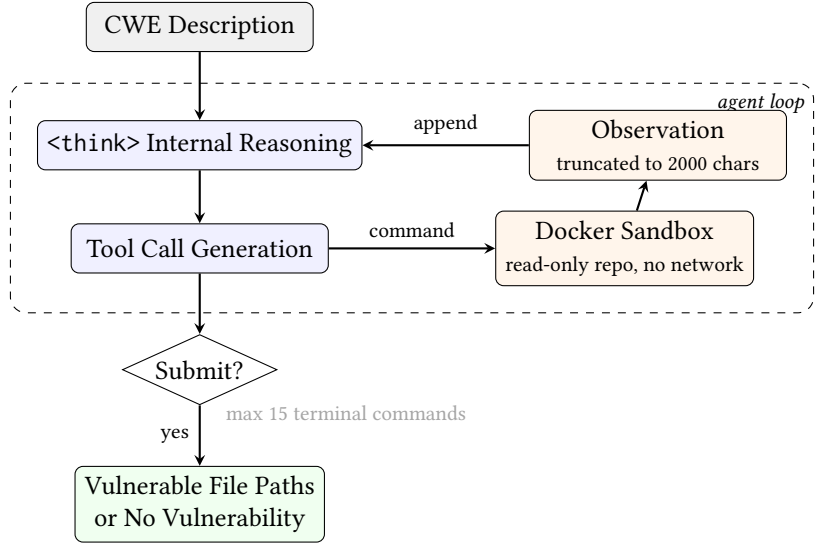
\begin{figure}[t]
\centering
\begin{tikzpicture}[
  node distance=0.9cm and 2.2cm,
  every node/.style={font=\small},
  box/.style={draw, rounded corners=3pt, minimum width=3cm, minimum height=0.65cm, align=center},
  input/.style={box, fill=gray!12},
  model/.style={box, fill=blue!6, minimum width=3.4cm},
  sandbox/.style={box, fill=orange!8, minimum width=3.4cm},
  output/.style={box, fill=green!6, minimum width=3cm},
  decision/.style={draw, diamond, aspect=2.2, inner sep=1.5pt, font=\small},
  arr/.style={->, thick, >=stealth},
  loopbox/.style={draw, rounded corners=6pt, dashed, inner sep=10pt}
]

\node[input] (input) {CWE Description};

\node[model, below=of input] (reason) {\texttt{<think>} Internal Reasoning};
\node[model, below=0.7cm of reason] (toolcall) {Tool Call Generation};

\node[decision, below=0.8cm of toolcall] (decision) {Submit?};

\node[sandbox, right=of toolcall] (sandbox) {Docker Sandbox\\{\scriptsize read-only repo, no network}};
\node[sandbox, right=of reason] (obs) {Observation\\{\scriptsize truncated to 2000 chars}};

\node[output, below=0.8cm of decision] (output) {Vulnerable File Paths\\or No Vulnerability};

\draw[arr] (input) -- (reason);
\draw[arr] (reason) -- (toolcall);
\draw[arr] (toolcall) -- (decision);
\draw[arr] (decision) -- node[left, font=\scriptsize] {yes} (output);
\draw[arr] (toolcall) -- node[above, font=\scriptsize] {command} (sandbox);
\draw[arr] (sandbox) -- (obs);
\draw[arr] (obs) -- node[above, font=\scriptsize] {append} (reason);

\node[font=\scriptsize, color=gray!70, below right=0.1cm and -0.3cm of decision] {max 15 terminal commands};

\begin{pgfonlayer}{background}
  \node[loopbox, fit=(reason)(toolcall)(sandbox)(obs), label={[font=\scriptsize\itshape, anchor=north east]north east:agent loop}] {};
\end{pgfonlayer}

\end{tikzpicture}
\caption{Antares inference loop for a single evaluation task. Given only a CWE category description, the model iteratively reasons, issues read-only terminal commands against the repository sandbox, and incorporates observations until it identifies vulnerable files or exhausts its turn budget.}
\label{fig:antares_pipeline}
\end{figure}

\subsection{Model Family}

Antares consists of three decoder-only transformer models containing 350M, 1B, and 3B parameters. All models are initialized from IBM Granite 4.0 checkpoints \cite{ibm2025granite4nano,ibm2025granite4_350m,ibm2025granite4_1b,ibm2025granite4_micro} and share the same tokenizer and architectural blueprint: grouped-query attention, SwiGLU MLP activations, RMSNorm, RoPE positional embeddings, and shared input/output embedding matrices. The three variants differ only in layer count, hidden dimension, and attention head configuration. Scaling the model family allows us to evaluate how repository exploration and vulnerability localization capabilities evolve with model capacity while maintaining a fixed training pipeline.

\begin{table}[h!]
\centering
\caption{Antares model variants with base checkpoint, maximum context length, and intended deployment tier. All three models target local, low-resource inference without requiring datacenter hardware.}
\label{tab:model_variants}
\begin{tabular}{lcccc}
\toprule
\textbf{Model} & \textbf{Parameters} & \textbf{Base Model} & \textbf{Context} & \textbf{Intended Deployment} \\
\midrule
Antares-350M & 350M & Granite 4.0 350M & 32K & Mobile / IoT \\
Antares-1B   & 1B   & Granite 4.0 1B   & 128K & Laptop / Workstation \\
Antares-3B   & 3B   & Granite 4.0 Micro & 128K & Laptop / Any Single GPU \\
\bottomrule
\end{tabular}
\end{table}

\subsection{Agentic Execution Environment}

Antares is not evaluated as a standalone sequence model. It operates inside a constrained agent loop that exposes a small set of read-only tools for repository exploration and final submission. Given a CWE category description, the model iteratively reasons, issues terminal commands, observes truncated command outputs, and either submits vulnerable file paths or declares that no vulnerability is present. The environment restricts interaction to read-only repository navigation and inspection, disables network access, and enforces a fixed terminal-command budget. These constraints make localization a controlled agentic task: the model must decide what to search, which files to inspect, when to refine its hypothesis, and when to submit.

This execution environment serves two roles. During training, it provides the multi-turn trajectories on which GRPO rewards are computed. During evaluation, it ensures that all models are compared under the same repository access, tool budget, and submission protocol. The Antares CLI described in Section~\ref{sec:cli} packages the Antares agent protocol for practical deployment. It preserves the benchmark's CWE-conditioned task, default repository-tool budget, and file-level submission semantics while adding production-specific repository isolation and integration features.

\subsection{Why Small Models?}
\label{subsec:whysmallmodels}
Cybersecurity workflows place different constraints on models than general coding benchmarks. In many real deployment settings, security tools need to run close to the codebase, integrate into existing developer and security workflows, and operate under strict privacy, latency, and cost constraints. This is especially important for vulnerability localization, where the model may need to inspect proprietary repositories and interact repeatedly with the environment before producing a result. The security domain further demands that inference remain in-house: sending source code to external APIs introduces supply-chain risk and is often prohibited by enterprise security policies.

Rather than relying exclusively on frontier-scale models, effective cybersecurity agents benefit from high tokens-per-second throughput at low cost. Multi-turn agentic loops amplify latency and cost linearly with conversation depth, making compact models that sustain high generation speed on commodity hardware particularly attractive for this setting. Antares-3B generates over 1,500 tokens per second on a single GPU, completing a full 500-task evaluation sweep spanning 290 repositories in $\sim$15 minutes. At average cloud rental rates (\$2--\$4 per H100-hour across commodity GPU providers), this translates to under one dollar per complete evaluation run, enabling repeated repository-scale evaluation at low marginal cost. Comparable frontier API usage exceeds one hundred dollars for the same workload.

We therefore design Antares around this assumption: a compact, specialized model trained directly for terminal-based vulnerability localization can deliver competitive accuracy while remaining practical for integration into CI/CD pipelines, code review systems, and security triage workflows that require fully local, closed-network operation.

%% file: content/4-data.tex
\section{Training Data}

Training Antares requires data that spans two complementary competencies, namely understanding vulnerability patterns and navigating unfamiliar codebases through terminal interaction. We construct separate datasets for supervised fine-tuning and reinforcement learning, reflecting the distinct objectives of each training stage. The SFT corpus establishes broad security knowledge, deep research capability, and terminal fluency, while the RL dataset supplies verifiable localization tasks constructed through proprietary curation pipelines.

\begin{table}[h!]
\centering
\caption{Training data composition across SFT and RL stages.}
\label{tab:data_stats}
\begin{tabular}{lcp{6.5cm}}
\toprule
\textbf{Component} & \textbf{Fraction} & \textbf{Purpose} \\
\midrule
\multicolumn{3}{l}{\textit{SFT Corpus}} \\
\quad Cybersecurity Reasoning & 71.5\% & Teaches vulnerability concepts, CWE/CVE reasoning, advisory interpretation, and security analysis. \\
\quad Deep Research \& General & 13.1\% & Preserves broad multi-step reasoning, evidence aggregation, and general instruction-following behavior. \\
\quad Code Search Trajectories & 15.4\% & Teaches terminal-based repository exploration, file inspection, and iterative code search. \\
\midrule
\multicolumn{3}{l}{\textit{RL Corpus}} \\
\quad Repository Localization Tasks & --- & Provides verifiable end-to-end vulnerability localization tasks over repository snapshots curated through proprietary pipelines. \\
\bottomrule
\end{tabular}
\end{table}

\subsection{Security and Deep Research Corpus}

The supervised fine-tuning dataset is organized into three categories that together cover the skills required for vulnerability localization. Following the methodology established in Foundation-Sec-8B-Reasoning \cite{yang2026foundationsec-reasoning}, SFT builds a broad foundation before reinforcement learning specializes the policy.

Cybersecurity reasoning and deep research data together use GPT-OSS-120B \cite{openai2025gptoss120bgptoss20bmodel} as the unified teacher model. This single-teacher design is deliberate: cross-domain teacher mixing has been shown to introduce distribution shifts and higher output entropy compared to single-teacher baselines \cite{falconllmteam2026falconh1rpushingreasoningfrontiers}, so we source all reasoning traces from one model to maintain consistent reasoning style across the entire non-terminal corpus. The security portion covers CWE taxonomy, CVE-to-CWE mappings, threat modeling, and advisory interpretation. The deep research component follows the long-horizon trajectory synthesis methodology of OpenResearcher \cite{li2026openresearcherfullyopenpipeline}, comprising multi-turn web search and evidence aggregation workflows that maintain broad multi-step reasoning capabilities.

\subsection{Terminal Trajectories}

Code search trajectories constitute the remaining 15\% of the SFT corpus. Each trace captures a complete tool-use conversation in which a model navigates a repository to locate specific files given natural-language descriptions. Traces follow the full interaction protocol consisting of system prompt, reasoning, terminal command, observation, and answer. Trajectories span Java, JavaScript, Go, Swift, Python, and additional ecosystems.

This component is critical because it teaches the model \emph{how} to explore repositories using only a terminal. Rather than relying on static retrieval, the model learns to issue appropriate commands, interpret directory structures, follow import chains, and refine searches based on intermediate observations. GRPO later specializes this general search capability toward vulnerability localization specifically.

All SFT data uses a unified message format with explicit reasoning traces wrapped in thinking tags. The cybersecurity reasoning component provides what to look for and the code search component teaches how to look, but these two capabilities remain disconnected until the reinforcement learning stage bridges them by optimizing for finding security vulnerabilities through terminal-based repository exploration.

\begin{figure}[t]
\centering
\begin{tikzpicture}

\slice{90}{1.8}{45.9}{1.55}{24.5\%}{fa1}
\slice{1.8}{-71.28}{-34.74}{1.55}{20.3\%}{fa2}
\slice{-71.28}{-123.84}{-97.56}{1.55}{14.6\%}{fa3}
\slice{-123.84}{-162.36}{-143.10}{1.55}{10.7\%}{fa4}
\slice{-162.36}{-199.80}{-181.08}{1.55}{10.4\%}{fa5}
\slice{-199.80}{-232.92}{-216.36}{1.55}{9.2\%}{fa6}

\slice{-232.92}{-251.28}{-242.10}{1.95}{5.1\%}{fa7}
\slice{-251.28}{-268.56}{-259.92}{2.05}{4.8\%}{fa8}

\filldraw[
  fill=fa9,
  draw=white,
  line width=1.2pt
]
(0,0) -- (-268.56:2.8)
arc[start angle=-268.56,end angle=-270,radius=2.8]
-- cycle;

\draw[->, thick]
  (0.08,2.78)
  to[out=95,in=210]
  (0.55,3.22);

\node[font=\small] at (0.78,3.34) {0.4\%};

\begin{scope}[shift={(3.7,1.6)}]
\node[anchor=west] at (0,0)       {\textcolor{fa1}{\rule{0.22cm}{0.22cm}} \quad pip (Python)};
\node[anchor=west] at (0,-0.38)   {\textcolor{fa2}{\rule{0.22cm}{0.22cm}} \quad npm (JavaScript)};
\node[anchor=west] at (0,-0.76)   {\textcolor{fa3}{\rule{0.22cm}{0.22cm}} \quad Go};
\node[anchor=west] at (0,-1.14)   {\textcolor{fa4}{\rule{0.22cm}{0.22cm}} \quad Maven (Java)};
\node[anchor=west] at (0,-1.52)   {\textcolor{fa5}{\rule{0.22cm}{0.22cm}} \quad Composer (PHP)};
\node[anchor=west] at (0,-1.90)   {\textcolor{fa6}{\rule{0.22cm}{0.22cm}} \quad Rust};
\node[anchor=west] at (0,-2.28)   {\textcolor{fa7}{\rule{0.22cm}{0.22cm}} \quad RubyGems};
\node[anchor=west] at (0,-2.66)   {\textcolor{fa8}{\rule{0.22cm}{0.22cm}} \quad NuGet (C\#)};
\node[anchor=west] at (0,-3.04)   {\textcolor{fa9}{\rule{0.22cm}{0.22cm}} \quad Other};
\end{scope}

\end{tikzpicture}

\caption{Ecosystem distribution of the RL training corpus. The dataset spans nine package ecosystems, with Python and JavaScript constituting the largest shares.}
\label{fig:grpo_ecosystem}
\end{figure}
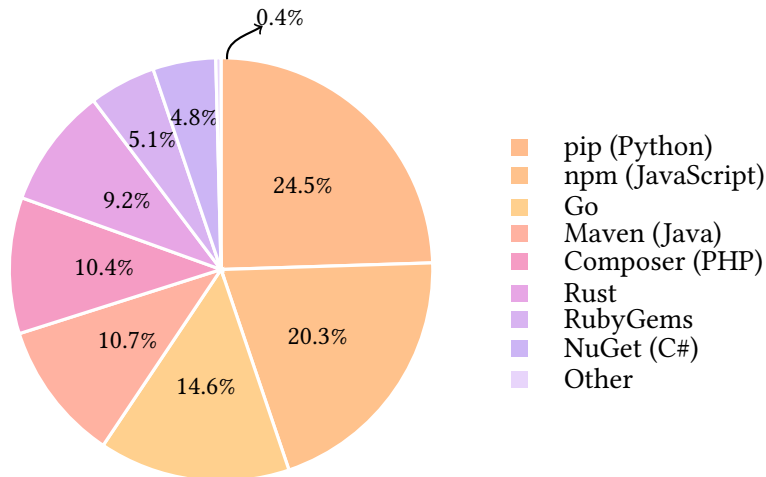

\subsection{Reinforcement Learning Dataset}

The reinforcement learning dataset consists of vulnerable repository snapshots curated through proprietary data-generation and filtering pipelines. Each task provides a complete codebase containing a known vulnerability alongside ground-truth file labels identifying the vulnerable source files.

The dataset covers 9 software ecosystems (pip, npm, Go, Maven, Composer, Rust, RubyGems, NuGet, and Others) spanning 255 unique CWE categories. During GRPO rollouts, each repository snapshot is extracted at runtime into an isolated Docker sandbox, providing the model with a realistic exploration environment identical in structure to production codebases.

Training and evaluation datasets are strictly disjoint. Throughout training, we verify that the GRPO training corpus has no overlap with \vlb{}, ensuring that performance reflects genuine generalization rather than memorization of specific repositories or vulnerability patterns.

\subsection{Data Filtering}

Ground-truth labels identify the implementation files containing each vulnerability. Test files, documentation, and configuration are excluded from the label set, ensuring that the reward signal during GRPO training reflects localization of vulnerable code rather than identification of ancillary changes.

%% file: content/5-training-pipeline.tex
\section{Training Pipeline}

Antares is trained using a two-stage post-training pipeline designed to progressively develop cybersecurity knowledge, terminal interaction skills, and repository exploration strategies. Rather than optimizing all capabilities simultaneously, each stage targets a distinct aspect of the vulnerability localization problem. The first stage (supervised fine-tuning) establishes format compliance, security domain knowledge, and basic terminal fluency. The second stage (reinforcement learning) optimizes the policy for end-to-end vulnerability localization over multi-turn agent trajectories.

\subsection{Supervised Fine-Tuning}

\subsubsection{Objective}

The SFT stage transforms each Granite 4.0 base model from a general-purpose language model into a terminal-capable security reasoning agent. The base models achieve near-zero File F1, on our evaluation benchmark prior to fine-tuning. While they possess tool-calling capabilities, they have no notion of structured terminal interaction and produce degenerate outputs when placed in an agentic loop. SFT addresses this by teaching three capabilities simultaneously: cybersecurity domain knowledge, structured deep-research behavior, and terminal interaction protocols.

\subsubsection{Training Procedure}

We fine-tune each model on the full SFT corpus for one epoch using AdamW ($\beta_1$=0.9, $\beta_2$=0.999), a learning rate of $5 \times 10^{-5}$ with cosine decay, and a global batch size matched to 8$\times$H100 throughput. Training completes on a single 8$\times$H100 node. All three model sizes (350M, 1B, 3B) use the same data mix and hyperparameters with no per-size tuning applied at this stage.

\subsubsection{Auxiliary Objectives}

Standard SFT trains the model to imitate assistant actions, including reasoning traces and tool calls, but it provides no direct supervision on how the model should represent environment observations. In our setting, these observations are central to the task: the model must interpret directory listings, search results, and file contents before deciding which command to issue next. Without an auxiliary signal, this grounding is learned only indirectly from the relationship between observations and subsequent actions, which is particularly challenging for smaller models.

We therefore incorporate auxiliary supervision during SFT to improve the model's representation of terminal feedback. The goal is to make environment observations useful for downstream repository navigation, while avoiding objectives that overfit to repository-specific surface forms.

A natural approach is token-level observation prediction. Objectives such as ECHO~\cite{shrivastava2026echo} add auxiliary cross-entropy loss on raw environment observation tokens, giving the model a direct learning signal on terminal outputs. This objective is well motivated: predicting observations can help the model learn how commands affect the environment and how terminal feedback should inform future actions. However, terminal outputs in repository exploration are highly instance-specific. Directory listings, file paths, and grep results vary substantially across codebases, so exact token prediction can emphasize surface-level reconstruction rather than transferable understanding of terminal feedback. We include this objective as an ablation in Section~\ref{sec:results}.

Antares instead uses semantic conditioning as its SFT auxiliary objective. Rather than predicting the exact tokens in an observation, semantic conditioning operates at the representation level. It encourages the model to learn similar internal representations for observations with similar functional roles, such as directory listings, search outputs, and source-code snippets, even when their surface tokens differ across repositories. This follows recent work on latent-space objectives for language models~\cite{jelassi2026matchingfeaturestokensenergybased, teoh2026nextlatentpredictiontransformerslearn}, which suggests that representation-level supervision can provide a more transferable learning signal when surface forms are highly variable.

As shown in Table~\ref{tab:aux_objectives}, semantic conditioning provides the strongest SFT initialization across all Antares model sizes, outperforming both standard SFT and token-level observation prediction. This stage establishes format compliance, security knowledge, terminal interaction, and observation grounding. Reinforcement learning is then used to optimize complete trajectories, converting this grounded terminal capability into a focused search--verify--refine policy for vulnerability localization.

\subsection{Reinforcement Learning}

\subsubsection{Objective}

The RL stage applies Group Relative Policy Optimization (GRPO) \cite{shao2024deepseekmath} to optimize the SFT policy for end-to-end vulnerability localization. Rather than using a learned reward model, we employ verifiable multi-component rewards computed programmatically from each trajectory. The objective is to transform unfocused terminal exploration into targeted, strategic vulnerability search.

\subsubsection{Environment}
\label{subsubsec:rl_environment}

During each GRPO rollout, a repository archive is extracted into an isolated,
Docker-backed workspace. The model interacts with the repository through the
same constrained interface used at evaluation time, with access limited to
read-only file-system navigation and inspection. Network access and package
installation are disabled, requiring the model to rely entirely on the
repository contents available within the workspace. Terminal command outputs
are truncated to 2,000 characters before being appended to the trajectory,
bounding context growth and preventing unusually long outputs from dominating
subsequent interactions.

\subsubsection{Agent Loop}
\label{subsubsec:agentloop}
Each rollout follows a fixed interaction protocol: the model receives a system prompt containing a CWE category description, then iteratively generates reasoning (wrapped in thinking tags), issues tool calls, receives observations, and continues until it submits a localization prediction or exhausts its turn budget. Three tools are available: \texttt{terminal} (read-only repository navigation and inspection commands), \texttt{submit\_vulnerable\_files} (ranked file paths), and \texttt{submit\_no\_vulnerability\_found} (declare clean). The training budget allows up to 15 assistant turns and 15 observation turns per rollout, followed by one final submission action. Rollout generation uses temperature 0.7, a maximum response length of 4,096 tokens, and a maximum model context of 16,384 tokens.

\subsubsection{GRPO Algorithm and Reward Function}

For each prompt, we sample multiple complete multi-turn trajectories, compute relative advantages from verifiable reward components, and update the policy using a clipped GRPO objective over assistant action tokens only.

We use a multi-component verifiable reward combining localization quality, valid submission behavior, tool-use compliance, exploration behavior, and penalties for malformed file predictions. All reward components are computed programmatically from trajectory text, with no learned reward model.

\begin{table}[h!]
\centering
\caption{GRPO reward components. The reward combines verifiable signals for localization accuracy, valid task completion, tool-use behavior, and malformed-output avoidance. All components are computed programmatically from trajectory text with no learned reward model.}
\label{tab:reward}
\begin{tabular}{lp{8.5cm}}
\toprule
\textbf{Component} & \textbf{Purpose} \\
\midrule
Localization quality & Measures agreement between submitted file paths and ground-truth vulnerable files. \\
\midrule
Submission behavior & Encourages the model to complete the task through the appropriate submission tools rather than failing to submit or prematurely declaring no vulnerability. \\
\midrule
Tool-use compliance & Rewards valid interaction with the agent loop and discourages malformed tool calls. \\
\midrule
Exploration behavior & Encourages the model to gather evidence from the repository before making a final localization prediction. \\
\midrule
Malformed-output penalty & Penalizes invalid or hallucinated file predictions that cannot be resolved cleanly through the structured submission interface. \\
\bottomrule
\end{tabular}
\end{table}

These components provide denser feedback than a binary success signal while preserving verifiability. Early in training, submission and tool-use signals help stabilize the agent loop and encourage consistent task completion. As the policy begins to submit valid predictions more reliably, localization quality becomes the primary signal distinguishing higher- and lower-quality trajectories within each group. Unlike Foundation-Sec-8B-Reasoning~\cite{yang2026foundationsec-reasoning}, which required an explicit format penalty to prevent reward hacking, our agent loop naturally constrains output format because the model must produce valid tool calls to receive observations.


\subsubsection{Infrastructure and Training Configuration}

We use veRL~\cite{sheng2025verl} as the training framework with vLLM as the inference backend. veRL was selected because its multi-turn rollout pipeline supports custom tool-call parser registration, enabling integration with Granite's tool-calling format for mid-trajectory generation and conversation management. Custom patches were required for Granite tool-call parser registration, domain-specific agent loop integration, and reward component logging.

Training uses a single 8$\times$ NVIDIA H100 80GB node. The actor model is distributed across GPUs with FSDP, while rollout inference uses vLLM with fixed rollout budgets across model scales. We use low-learning-rate GRPO with KL regularization against the SFT reference policy, small rollout groups, and optimizer states offloaded to CPU.

\subsection{Discussion}

The most visible effect of GRPO is reduced performance variance across rollouts. The SFT policy exhibits high stochasticity with inconsistent trajectories across different repositories. After GRPO, the policy produces stable and repeatable search strategies, transitioning from unfocused exploration to targeted vulnerability search given a particular CWE category and repository structure. Quantitative results across model scales are presented in Section~\ref{sec:results}.

%% file: content/6-inference-pipeline.tex
\section{Deployment: Antares CLI}
\label{sec:cli}
The Antares CLI packages the agent protocol described in
Section~\ref{subsubsec:agentloop} as a deployment interface for file-level
vulnerability localization. It preserves the core evaluation semantics: the
model receives a CWE-conditioned prompt, explores the repository under a
bounded inspection budget, and either submits file paths or declares
that no matching vulnerability was found. The default 15-call inspection
budget matches the evaluation configuration but can be adjusted at deployment.
The production harness additionally operates over an immutable repository
snapshot and provides a dedicated file-reading tool.

The CLI supports both targeted analyses over explicit CWE identifiers and
repository-wide sweeps over user-specified or automatically selected CWE sets.
For automatic selection, the repository is profiled against the bundled MITRE
CWE taxonomy before independent investigations are launched in parallel. A
local planning mode allows users to preview the selected categories and
supporting evidence without invoking the model. The CLI produces both
human-readable and structured reports, including SARIF~2.1.0 output for GitHub
Code Scanning. An optional failure-on-findings policy supports CI gating
without treating candidate detections as fatal by default. Outputs are
file-level candidates intended for human validation; line-level localization and remediation remain outside the current system's scope.

A non-interactive JSON interface supports integration with coding assistants
and orchestration frameworks by accepting structured requests and returning
findings, summary statistics, metadata, and per-CWE results. This interface
allows external agents to invoke Antares as a tool without requiring PTY or
signal management. The CLI delegates model execution to a user-configured,
streaming, OpenAI-compatible endpoint rather than hosting inference directly.
The released Antares-350M and Antares-1B models can be served within the user's
environment, enabling closed-network or air-gapped operation when both the
model weights and inference endpoint are hosted locally.

%% file: content/7-evaluation.tex
\section{Experimental Setting}
\label{sec:experimental_setting}

\subsection{Benchmark}

We evaluate all models on \vlb{}, a vulnerability-localization benchmark comprising 500 tasks drawn from 290 unique real-world repositories, spanning 6 package ecosystems and 147 unique CWE categories, with 78\% of entries carrying assigned CVE identifiers. Some repositories contribute multiple tasks corresponding to distinct vulnerabilities, advisories, or pull requests. Each task pairs a repository snapshot containing a known vulnerability with ground-truth implementation file labels. Evaluation proceeds in two phases: Phase A (localization) requires identifying vulnerable files given a CWE description, while Phase B (verification) presents patched code and expects the model to declare no vulnerability present. The experiments in this report focus on Phase A localization; Phase B is
included in the benchmark specification but is not evaluated here.

\subsection{Metrics}

We evaluate whether models identify the correct vulnerable files, rather than simply classifying a repository as vulnerable. For each task, we compare the submitted files with the ground-truth set and compute file-level precision, recall, and F1. We macro-average each metric across all 500 tasks. 

\begin{itemize}
\item \textbf{File F1:} For each task, the harmonic mean of file-level precision and recall. This is the primary metric used for model comparisons.
\item \textbf{Precision:} For each task, the fraction of submitted file paths that appear in the ground-truth set.
\item \textbf{Recall:} For each task, the fraction of ground-truth vulnerable files included in the submitted file paths.
\item \textbf{Abstain Rate:} The fraction of tasks for which the model submits no vulnerable file paths, either by explicitly declaring that no matching vulnerability was found or by failing to produce a valid localization submission.
\end{itemize}

Because every Phase A task contains a known vulnerability, abstaining receives zero precision, recall, and File F1 for that task.

\subsection{Models Evaluated}

We compare Antares against frontier closed-source models, large open-weight models, and small open-weight models, all evaluated using the same harness, tools, task inputs, interaction budget, and generation settings of temperature 0.3 and top-$p$ 1.0. We run each model three times and report the average across the three runs.

\begin{itemize}
\item \textbf{Frontier (closed):} GPT-5.5 (reasoning\_effort\,=\,\texttt{default}, \texttt{xhigh}), GPT-5, GPT-5 Mini, GPT-5 Nano, Gemini 3 Pro, Gemini 2.5 Flash, Gemini 3.1 Flash Lite.
\item \textbf{Open-weight large ($\geq$20B):} GLM-5.2, MiniMax-M2.7, Qwen3.5-122B-A10B, GPT-OSS-120B, Llama-3.3-70B, Qwen3.5-35B-A3B, Gemma-4-31B, Qwen3.5-27B, GPT-OSS-20B.
\item \textbf{Open-weight small ($<$20B):} CodeScout-14B, Qwen3.5-9B, Gemma-4-E4B, Gemma-4-E2B.
\item \textbf{Antares family:} 350M (SFT, GRPO), 1B (SFT, GRPO), 3B (SFT, GRPO).
\item \textbf{Baselines:} Granite 4.0 base models (350M, 1B, 3B) without any post-training.
\end{itemize}

\subsection{Evaluation Protocol}

All models are evaluated on the benchmark using an identical agent protocol. Each task runs in a fresh Docker container (Ubuntu 24.04) with 2 CPU cores, 4GB RAM, network disabled, and a 10-second per-command timeout. The container is destroyed after each task.

The agent protocol provides a budget of 15 terminal commands per task, followed by one final submission action, with up to 3 retries on unsuccessful commands. The submission action does not count toward the terminal-call budget. Three tools are available: \texttt{terminal} (read-only repository navigation and inspection commands), \texttt{submit\_vulnerable\_files} (ranked file paths), and \texttt{submit\_no\_vulnerability\_found} (declare clean). The model receives only the CWE category description as input, with no advisory text, file hints, or severity details.

Antares models are served via vLLM on a single GPU (bfloat16, max-model-len 32768) with temperature 0.3 and top-p 1.0. External models use their respective API endpoints. Evaluation runs 16 parallel workers processing entries concurrently.

%% file: content/8-results.tex
\section{Results}
\label{sec:results}

We evaluate Antares across model scale, vulnerability structure, repository complexity, agent behavior, and training stage. Across these analyses, a consistent picture emerges: repository-scale vulnerability localization is not primarily a general-purpose code understanding benchmark. It rewards models that learn how to search, verify, and submit under a constrained interaction budget.

\subsection{Task-Specific Training Dominates Parameter Scale}
\label{subsec:main_performance}

\begin{table}[t]
\centering
\caption{Frontier model comparison on \vlb{} (Phase A). Performance exhibits a capability cliff: models either achieve the 0.186--0.229 tier or fall below 0.152 regardless of general-purpose scale.}
\label{tab:main_results}
\renewcommand{\arraystretch}{1.18}    
\setlength{\tabcolsep}{8pt}           
\begin{tabular}{lccc}
\toprule
\textbf{Model} & \textbf{File F1} & \textbf{Precision} & \textbf{Recall} \\
\midrule
GPT-5.5 (xhigh) & 0.229 & 0.310 & 0.221 \\
\addlinespace[2pt]
\rowcolor{antares}
\textbf{Antares-3B} & \textbf{0.223} & \textbf{0.303} & \textbf{0.221} \\
\addlinespace[2pt]
GPT-5.5 (default) & 0.221 & 0.305 & 0.211 \\
\addlinespace[2pt]
\rowcolor{antares}
\textbf{Antares-1B} & \textbf{0.209} & \textbf{0.262} & \textbf{0.224} \\
\addlinespace[2pt]
GLM-5.2 & 0.186 & 0.226 & 0.186 \\
Gemini 3 Pro & 0.152 & 0.190 & 0.153 \\
\addlinespace[2pt]
\rowcolor{antares}
\textbf{Antares-350M} & \textbf{0.135} & \textbf{0.136} & \textbf{0.178} \\
\addlinespace[2pt]
Gemini 2.5 Flash & 0.102 & 0.132 & 0.098 \\
GPT-5 Mini & 0.098 & 0.115 & 0.096 \\
Gemini 3.1 Flash Lite & 0.095 & 0.131 & 0.090 \\
GPT-5 & 0.048 & 0.062 & 0.048 \\
GPT-5 Nano & 0.024 & 0.038 & 0.021 \\
\bottomrule
\end{tabular}
\end{table}

\begin{table}[t]
\centering
\caption{Comparison of open-weight models and static analysis tools on \vlb{}. Antares achieves the strongest localization performance, suggesting that the task requires capabilities beyond general-purpose scale and rule-based analysis.}
\label{tab:open_models}
\renewcommand{\arraystretch}{1.18}    
\setlength{\tabcolsep}{8pt}           
\begin{tabular}{lrccc}
\toprule
\textbf{Model} & \textbf{Params} & \textbf{File F1} & \textbf{Precision} & \textbf{Recall} \\
\midrule
\multicolumn{5}{c}{\textit{Open-Weight Models}} \\
\midrule
\midrule

\addlinespace[2pt]
\rowcolor{antares}
\textbf{Antares-3B} & \textbf{3B} & \textbf{0.223} & \textbf{0.303} & \textbf{0.221} \\
\addlinespace[2pt]

\addlinespace[2pt]
\rowcolor{antares}
\textbf{Antares-1B} & \textbf{1B} & \textbf{0.209} & \textbf{0.262} & \textbf{0.224} \\
\addlinespace[2pt]

GLM-5.2 & 753B & 0.186 & 0.226 & 0.186 \\
\addlinespace[2pt]
\rowcolor{antares}
\textbf{Antares-350M} & \textbf{350M} & \textbf{0.135} & \textbf{0.136} & \textbf{0.178} \\
\addlinespace[2pt]

Gemma-4-31B & 31B & 0.101 & 0.131 & 0.097 \\
Qwen3.5-27B & 27B & 0.091 & 0.116 & 0.088 \\
Qwen3.5-122B-A10B & 125B & 0.091 & 0.124 & 0.083 \\
Qwen3.5-35B-A3B & 36B & 0.085 & 0.115 & 0.081 \\
GPT-OSS-20B & 20B & 0.070 & 0.095 & 0.065 \\
GPT-OSS-120B & 120B & 0.069 & 0.095 & 0.062 \\
MiniMax-M2.7 & 229B & 0.054 & 0.078 & 0.050 \\
CodeScout-14B & 14B & 0.044 & 0.065 & 0.039 \\
Qwen3.5-9B & 9B & 0.043 & 0.058 & 0.039 \\
Gemma-4-E2B & 2B & 0.039 & 0.045 & 0.042 \\
Gemma-4-E4B & 4B & 0.034 & 0.039 & 0.034 \\
Llama-3.3-70B & 70B & 0.012 & 0.016 & 0.014 \\

\midrule
\multicolumn{5}{c}{\textit{Static Analysis Tools}} \\
\midrule\midrule
Semgrep & N/A & 0.086 & 0.091 & 0.155 \\
Semgrep-CWE & N/A & 0.052 & 0.057 & 0.071 \\
CodeQL & N/A & 0.023 & 0.025 & 0.030 \\
 Horusec & N/A & 0.020 & 0.021 & 0.038 \\

\bottomrule
\end{tabular}
\end{table}

Tables~\ref{tab:main_results} and~\ref{tab:open_models} evaluate whether vulnerability localization improves smoothly with model scale. If general-purpose scale were sufficient, we would expect larger open-weight and frontier models to dominate smaller specialized models. Instead, the results show a capability cliff. GPT-5.5, Antares-3B, and GLM-5.2 form the only high-performing tier, while many larger general-purpose models fall far below this range.

Antares-3B reaches 0.223 File F1, approaching GPT-5.5 while outperforming substantially larger open-weight models, including GLM-5.2. The comparison against static analysis tools shows the same pattern from the opposite direction: rule-based scanners recover some vulnerable files, but remain below Antares models because they lack the ability to adaptively inspect repository context. Parameter count alone is therefore a poor predictor of localization performance.

The precision--recall decomposition further suggests that different model sizes learn different operating regimes. Antares-3B behaves conservatively, matching GPT-5.5's recall while maintaining high precision. Antares-1B achieves the highest recall of all evaluated systems, suggesting a search-heavy strategy that finds more candidate vulnerable files. Antares-350M shifts further toward recall at lower precision, consistent with a smaller model that can search effectively but has weaker verification capacity.

\begin{observationbox}[title={Observation 1}]Repository-scale vulnerability localization exhibits a capability cliff rather than smooth scaling. Compact models trained specifically for agentic vulnerability localization can outperform open-weight models hundreds of times larger, indicating that task-specific interaction training matters more than parameter count alone.
\end{observationbox}

\subsection{Localization Difficulty Follows Structure, Not Severity}
\label{subsec:dimensional_analysis}

\begin{figure}[t]
\centering
    \includegraphics[width=\linewidth]{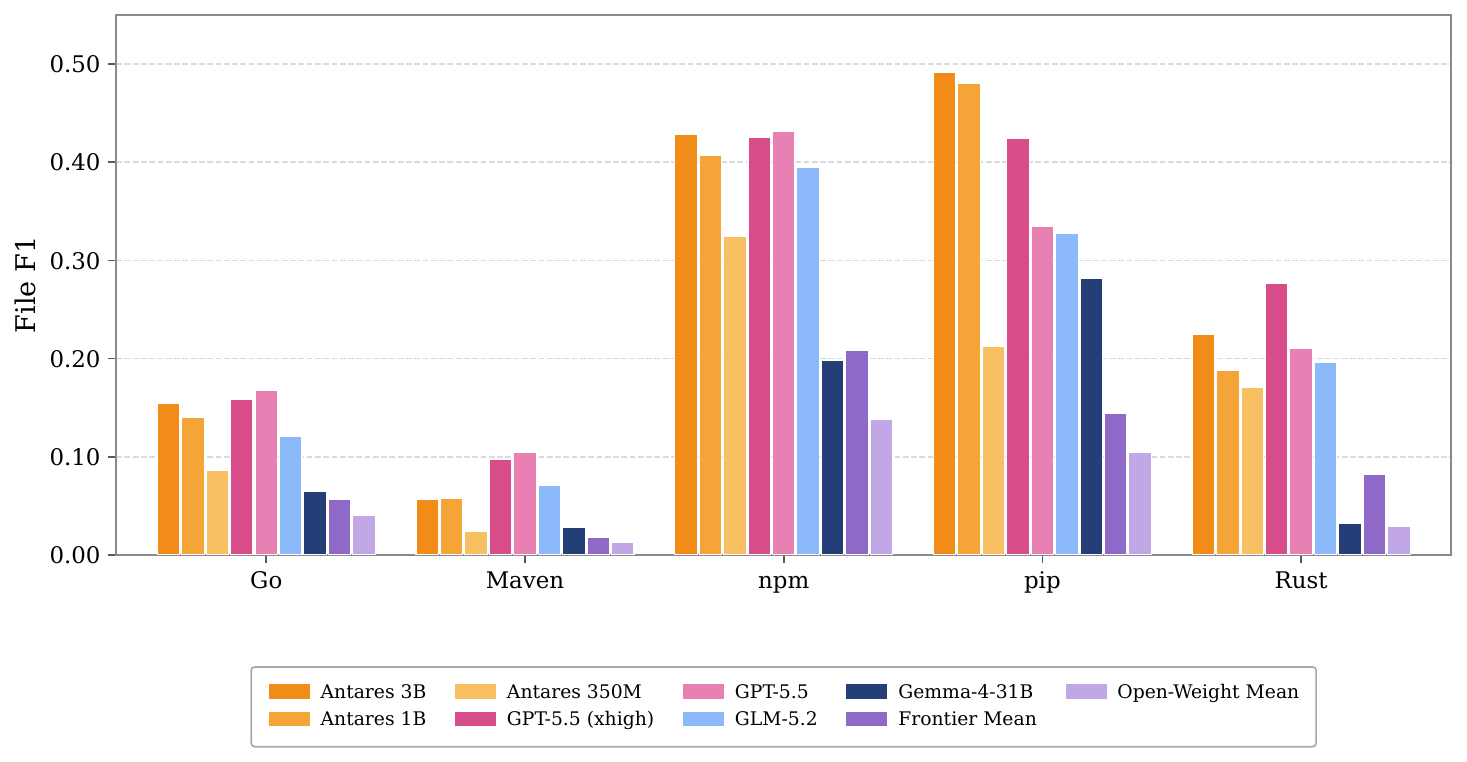}
\caption{File F1 disaggregated by package ecosystem (top 5 by frequency). Task counts: Go (n=215), Maven (n=104), npm (n=88), pip (n=52), Rust (n=40). Ecosystem structure determines difficulty uniformly across all models, with pip and npm yielding 7--14$\times$ higher scores than Maven regardless of model scale.}
\label{fig:perf_by_ecosystem}
\end{figure}

\begin{figure}[t]
\centering
    \includegraphics[width=\linewidth]{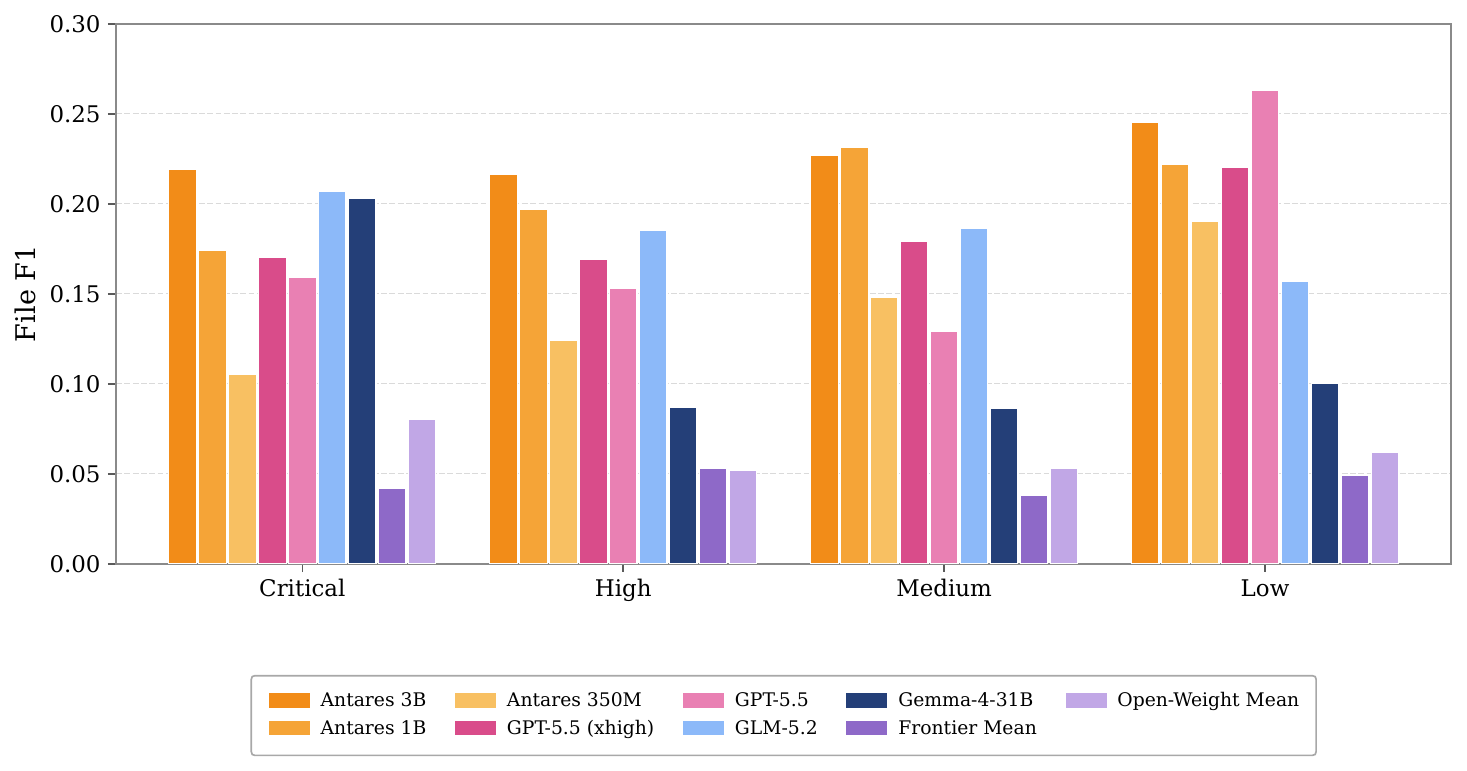}
\caption{File F1 disaggregated by CVSS severity. Task counts: Critical (n=57), High (n=219), Medium (n=194), Low (n=30). Performance varies less across severity bins than across ecosystem and
repository-structure categories, confirming that localization difficulty is determined by repository structure rather than vulnerability impact.}
\label{fig:perf_by_severity}
\end{figure}

We next ask what makes a vulnerability localization instance difficult. A natural hypothesis is that more severe vulnerabilities should be easier to find because they may correspond to more obvious or security-critical code. The results do not support this hypothesis. Instead, difficulty is dominated by repository and vulnerability structure.

Figure~\ref{fig:perf_by_ecosystem} shows that ecosystem structure strongly affects localization performance. Flat, convention-heavy ecosystems such as pip and npm produce the highest scores across models, while Maven remains difficult for every system. This suggests that localization is easier when vulnerable logic is concentrated in shallow, predictable paths, and harder when evidence is distributed across verbose build hierarchies, framework conventions, and multi-class implementations.

Figure~\ref{fig:perf_by_severity} shows the opposite pattern for CVSS severity. Performance varies little across Critical, High, Medium, and Low bins, indicating that vulnerability impact is not the main driver of localization difficulty. The CWE-level results reinforce this interpretation: structurally distinctive vulnerabilities such as code injection and deserialization are easier for Antares, while diffuse data-flow categories such as information exposure remain difficult for every model.

\begin{observationbox}[title={Observation 2}]Localization difficulty is governed by structural locality rather than vulnerability severity. Ecosystem conventions and CWE-specific implementation patterns determine whether an agent can efficiently narrow the search space.
\end{observationbox}

\subsection{Repository Scale and Multi-File Vulnerabilities Remain the Core Bottleneck}
\label{subsec:repo_complexity}

\begin{figure}[t]
\centering
    \includegraphics[width=\linewidth]{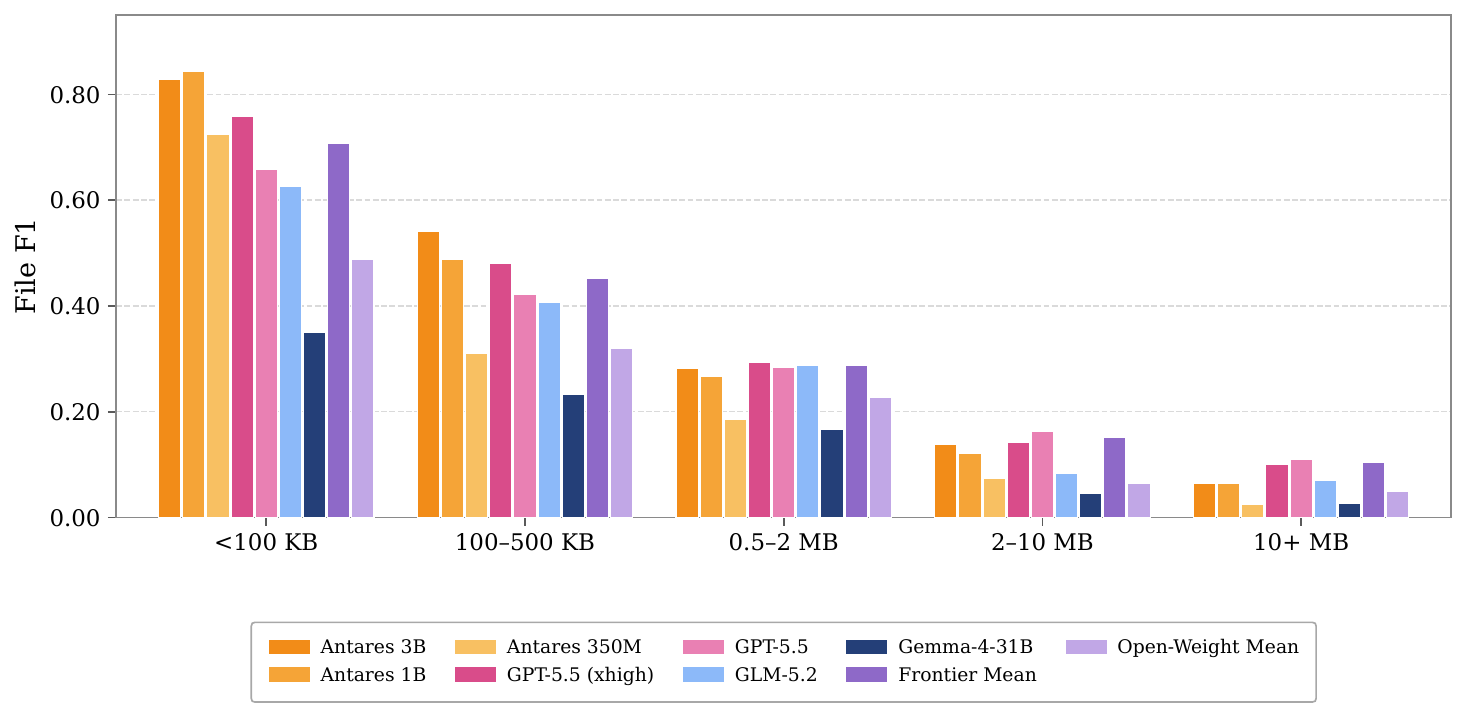}
\caption{File F1 by repository size (total codebase). Task counts: {<}100 KB (n=20), 100--500 KB (n=82), 0.5--2 MB (n=84), 2--10 MB (n=91), 10+ MB (n=223). All models decline sharply as repository size increases, but Antares maintains competitive or superior performance at every scale.}
\label{fig:repo_size}
\end{figure}

We then examine whether repository complexity changes which model strategy is most effective. Our hypothesis is that RL-trained search policies should excel when the repository is small enough for terminal exploration to cover most relevant files, while larger repositories should favor models with stronger long-horizon reasoning.

Figure~\ref{fig:repo_size} supports this hypothesis. On repositories under 100 KB, Antares models achieve the strongest performance of any evaluated system, with Antares-1B reaching 0.843 File F1 and Antares-3B reaching 0.828. In this regime, grep-based elimination and targeted file inspection are sufficient to approach complete coverage within the turn budget. As repositories grow, however, the advantage narrows. At the 10+ MB tier, GPT-5.5 variants overtake Antares, suggesting that larger repositories require architectural reasoning beyond efficient search.

The number of ground-truth files provides a second complexity axis. Single-file vulnerabilities are substantially easier for all models, while entries with five or more vulnerable files cause performance to collapse. This shows that current agents are much better at identifying a vulnerable foothold than recovering every file involved in a distributed vulnerability.

\begin{observationbox}[title={Observation 3}]RL-trained search policies are most effective when repository evidence can be covered within the interaction budget. Performance degrades sharply as vulnerabilities become distributed across larger codebases and multiple ground-truth files.
\end{observationbox}

\subsection{GRPO Induces a Search--Verify--Refine Policy}
\label{subsec:behavioral_strategy}

\begin{figure}[t]
\centering
\includegraphics[width=\linewidth]{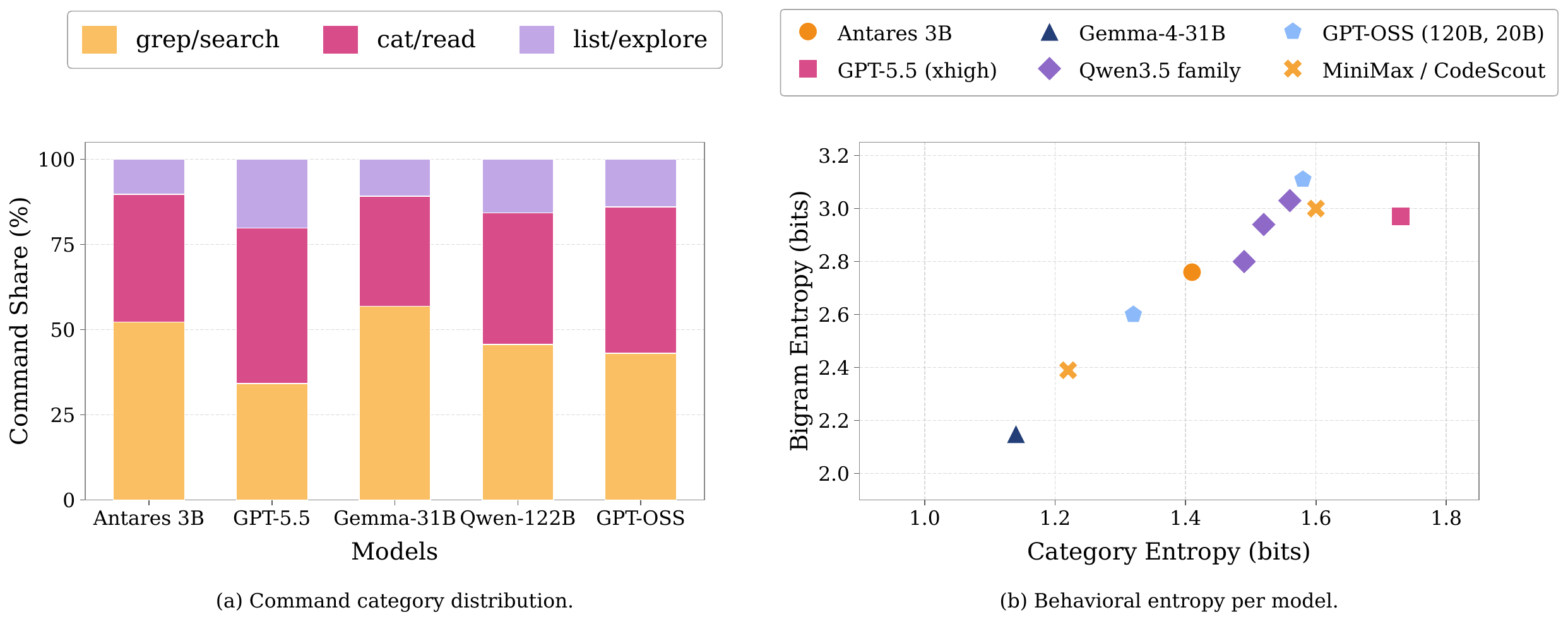}
\caption{Behavioral strategy analysis across 500 tasks from one representative run. (a) Command allocation shows Antares-3B maintains the highest search ratio among models with balanced read profiles. (b) Entropy scatter reveals Antares-3B achieves lower category entropy than all frontier models while sustaining moderate transition complexity through its search-verify loop.}
\label{fig:behavior_combined}
\end{figure}

The aggregate scores show that Antares is competitive, but they do not explain how it behaves. We therefore ask whether Antares solves localization by imitating frontier-style repository comprehension, or whether GRPO induces a distinct search policy.

Figure~\ref{fig:behavior_combined} shows that Antares-3B uses a narrower command repertoire than frontier models while maintaining non-trivial transition complexity. It relies heavily on search commands, uses fewer structural exploration commands, and issues fewer total commands per task than GPT-5.5. This suggests that Antares does not try to build a complete mental model of the repository. Instead, it treats localization as elimination: search broadly for vulnerability-relevant terms, read candidate files, and refine the search based on evidence.

The entropy analysis clarifies that this behavior is not simply rigid repetition. Antares has lower category entropy than frontier models, but its bigram entropy remains close to GPT-5.5 and GPT-OSS. In other words, Antares uses fewer action types, but it adapts how it transitions between them. Its behavioral complexity is concentrated in the search--verify--refine loop, which is directly aligned with the file-localization objective.

\begin{observationbox}[title={Observation 4}]GRPO induces a specialized search--verify--refine policy. Antares-3B is less behaviorally broad than frontier models, but its transitions remain adaptive and concentrated on the actions most useful for file-level localization.
\end{observationbox}

\subsection{Training Progression and Emergent Specialization}
\label{sec:training_progression}

We next ask how each stage of the Antares training pipeline contributes to the final agent. The central hypothesis is that SFT and GRPO play different roles: SFT should teach the model how to operate in the terminal environment, while GRPO should teach the model which interaction strategies are useful for vulnerability localization. We test this by comparing base, SFT, and GRPO checkpoints across all three model scales, then analyzing how behavior changes after reinforcement learning.

\begin{figure}[t]
\centering
\includegraphics[width=\linewidth]{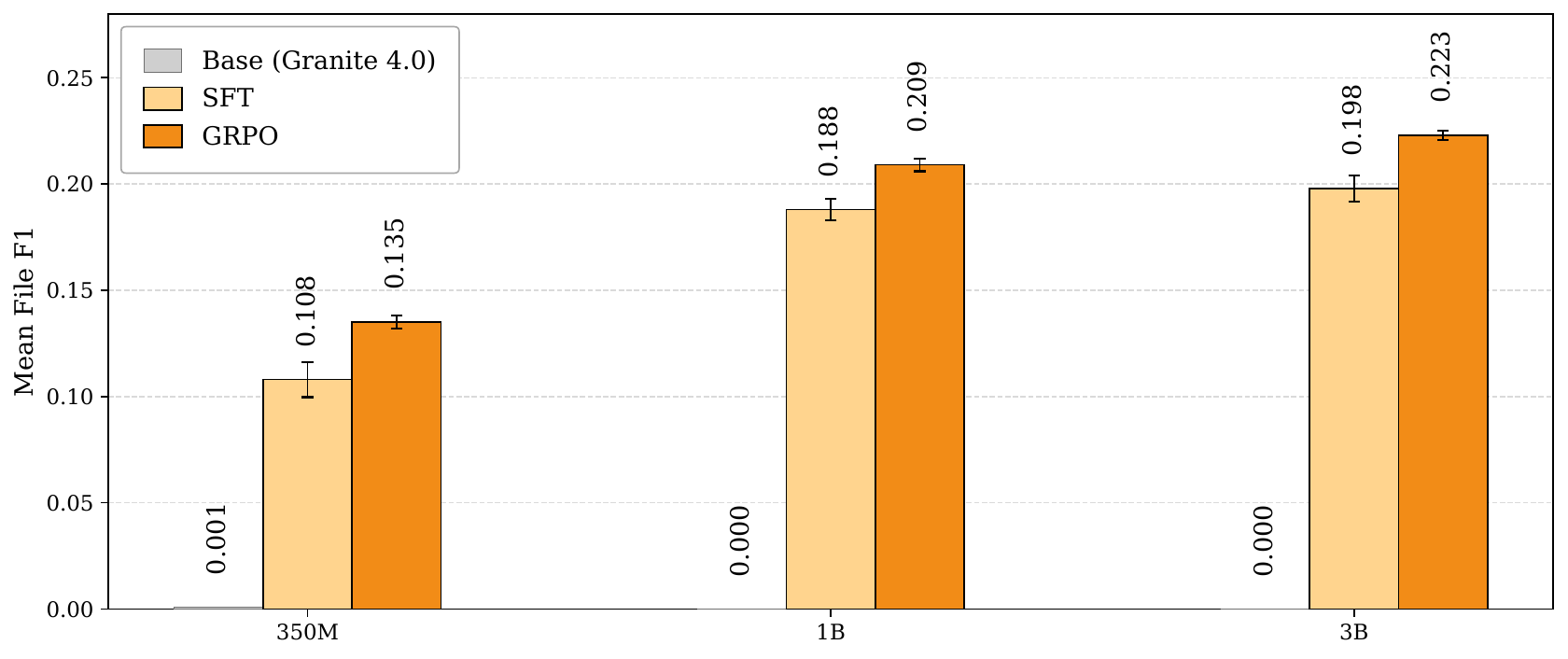}
\caption{Mean File F1 across the three-stage training pipeline (Base $\rightarrow$ SFT $\rightarrow$ GRPO) for each Antares model scale, averaged over 3 independent evaluation runs on \vlb{}.}
\label{fig:training_progression}
\end{figure}

\begin{table}[t]
\centering
\caption{\textbf{Operational metrics from SFT to GRPO across model scales.}
Values are averaged across three runs. $\sigma$ denotes the standard deviation of File F1 across runs, and Files Sub. denotes the mean number of file paths submitted per task.}
\label{tab:sft_grpo_behavioral}
\renewcommand{\arraystretch}{1.18}    
\setlength{\tabcolsep}{8pt}           
\begin{tabular}{llcccccc}
\toprule
\textbf{Scale} & \textbf{Stage} & \textbf{File F1} & \textbf{$\sigma$} & \textbf{Prec.} & \textbf{Rec.} & \textbf{Abstain} & \textbf{Files Sub.} \\
\midrule
\addlinespace[2pt]
\rowcolor{antares-sft} 350M & SFT & 0.108 & 0.0083 & 0.149 & 0.101 & 5.8\% & 1.30 \\
\addlinespace[2pt]
\rowcolor{antares} \textbf{350M} & \textbf{GRPO} & \textbf{0.135} & \textbf{0.0031} & \textbf{0.136} & \textbf{0.178} & \textbf{1.4\%} & \textbf{4.23} \\
\addlinespace[2pt]
\midrule
\addlinespace[2pt]
\rowcolor{antares-sft} 1B & SFT & 0.188 & 0.0052 & 0.263 & 0.179 & 3.0\% & 1.55 \\
\addlinespace[2pt]
\rowcolor{antares} \textbf{1B} & \textbf{GRPO} & \textbf{0.209} & \textbf{0.0030} & \textbf{0.262} & \textbf{0.224} & \textbf{0.6\%} & \textbf{2.95} \\
\addlinespace[2pt]
\midrule
\addlinespace[2pt]
\rowcolor{antares-sft} 3B & SFT & 0.198 & 0.0062 & 0.240 & 0.228 & 7.5\% & 2.68 \\
\addlinespace[2pt]
\rowcolor{antares} \textbf{3B} & \textbf{GRPO} & \textbf{0.223} & \textbf{0.0022} & \textbf{0.303} & \textbf{0.221} & \textbf{4.0\%} & \textbf{1.84} \\
\addlinespace[2pt]
\bottomrule
\end{tabular}
\end{table}

\paragraph{Performance Progression}

Figure~\ref{fig:training_progression} quantifies the contribution of each training stage. The base Granite models achieve near-zero File F1, indicating that general tool-calling ability alone is insufficient for repository-scale vulnerability localization. SFT with semantic conditioning lifts all three model scales into functional localization agents, with the 1B and 3B models clustering together at 0.188 and 0.198 File F1, while the 350M model remains lower at 0.108. This suggests that terminal-based vulnerability localization requires a minimum capacity threshold, but that SFT alone can already teach the basic interaction protocol.

GRPO then improves all three scales, but the gains are not uniform. The 350M model receives the largest relative improvement, increasing by 25\%, while the 1B and 3B models improve by 11--13\%. This pattern suggests that GRPO is most valuable when the SFT policy has learned the environment format but has not yet discovered reliable search behavior. At larger scales, the SFT initialization already captures more of the useful strategy space, leaving less room for policy optimization to improve mean performance.

The error bars in Figure~\ref{fig:training_progression} are as important as the mean improvements. GRPO reduces run-to-run standard deviation by 42--65\% across all scales (Table~\ref{tab:sft_grpo_behavioral}), indicating that RL collapses the policy toward a smaller set of high-return trajectories rather than merely increasing average performance. This variance reduction is operationally important: a single GRPO evaluation run provides a more reliable estimate of model behavior than a single SFT run.

\begin{observationbox}[title={Observation 5}]
SFT teaches Antares to operate as a terminal agent with security knowledge, while GRPO makes that behavior more reliable and task-directed. The primary effect of GRPO is not only higher File F1, but lower variance and more stable localization behavior.
\end{observationbox}

\paragraph{Emergent Scale-Dependent Specialization}

We then ask whether GRPO induces the same strategy across model scales. Before RL, all three models follow a similar read-dominant imitation policy, with 27--32\% search commands and 54--60\% file-reading commands. This is expected: SFT trains all scales on the same behavioral distribution, so the models initially imitate the same style of repository exploration.

After GRPO, this shared behavior disappears. Because the reward does not prescribe a specific distribution over search, read, and exploration command categories, each model scale discovers its own operating point. The 350M and 1B models shift sharply toward search-heavy behavior, using 87--89\% search commands and submitting more files. This high-recall strategy compensates for limited verification capacity by maximizing coverage. In contrast, the 3B model maintains a more balanced search/read policy, using 52\% search and 37\% read commands, while submitting fewer files at higher precision. This divergence suggests that GRPO does not teach a single universal localization algorithm. Instead, it exposes a capacity-dependent tradeoff between search coverage and verification quality. Smaller models benefit from broad search and higher submission volume, while the 3B model has enough capacity to verify candidates more selectively.

\begin{observationbox}[title={Observation 6}]GRPO induces scale-dependent specialization. Smaller Antares models learn high-recall search-heavy policies, while Antares-3B learns a more selective search-and-verify policy with higher precision.
\end{observationbox}

\paragraph{Auxiliary Objective Ablation}

\begin{table}[h!]
\centering
\caption{SFT File F1 by auxiliary objective. Semantic conditioning produces the strongest initialization for GRPO at all scales, with the largest margin at 350M where explicit terminal grounding is most critical.}
\label{tab:aux_objectives}
\renewcommand{\arraystretch}{1.18}
\setlength{\tabcolsep}{8pt}
\begin{tabular}{lccc}
\toprule
\textbf{Auxiliary Objective} & \textbf{350M} & \textbf{1B} & \textbf{3B} \\
\midrule
No auxiliary loss & 0.021 & 0.151 & 0.164 \\
ECHO \cite{shrivastava2026echo} & 0.076 & 0.174 & 0.177 \\
\addlinespace[2pt]
\rowcolor{antares} \textbf{Semantic conditioning (ours)} & \textbf{0.108} & \textbf{0.188} & \textbf{0.198} \\
\addlinespace[2pt]
\bottomrule
\end{tabular}
\end{table}

Finally, we test whether the SFT auxiliary objective affects downstream agent quality. Table~\ref{tab:aux_objectives} shows that semantic conditioning is the strongest initialization at every model scale. The effect is largest at 350M, where semantic conditioning improves File F1 by 5.1$\times$ over no auxiliary loss. At 3B, the gain narrows to 20.7\%, suggesting that larger models can partially infer terminal dynamics from ordinary next-token supervision, while compact models require more explicit grounding.

This result supports the view that auxiliary objectives shape the behavioral distribution available to GRPO. A weak SFT initialization gives RL fewer useful trajectories to reinforce; a stronger initialization exposes more viable search, read, and submit behaviors. Semantic conditioning therefore acts less like a small additive improvement and more like a multiplier on the effectiveness of the entire post-training pipeline.

\begin{observationbox}[title={Observation 7}]
    Semantic conditioning provides the strongest SFT initialization for terminal-based vulnerability localization. Its effect is largest for compact models, where explicit grounding of terminal observations is most important.
\end{observationbox}

\subsection{Does Vulnerability-Localization Training Transfer Beyond \vlb{}?}
\label{sec:generalization_beyond_vloc}

The evaluations above focus on repository-scale vulnerability localization, the
capability directly optimized during Antares training. To test whether the
learned policy transfers beyond this setting, we additionally evaluate Antares
on issue-driven code localization and structured multi-turn tool use. Full
results are reported in \autoref{app:additional_benchmarks}.

On SWE-Bench, Antares remains competitive with substantially larger models
trained specifically for code localization, despite having no exposure to
SWE-Bench repositories, issue descriptions, or issue-resolution data. On
BFCL-v3, Antares shows its largest gains over the corresponding Granite base
models in multi-turn orchestration, while aggregate function-calling
performance remains broadly comparable. Together, these results indicate that
Antares acquires transferable repository-navigation and sequential tool-use
capabilities rather than a policy narrowly specialized to \vlb{}.

%% file: content/9-error_analysis.tex
\section{Discussion}
\label{sec:error_analysis}

\paragraph{Where Current Agents Still Fail}
\label{subsec:failure_modes}

The hardest entries are not random failures; they concentrate in repositories with large search spaces, many files, and diffuse vulnerability evidence. In the highest-complexity quartile, Antares-3B falls below 0.04 File F1, compared to 0.55 on the easiest quartile. This degradation is shared across models: even GPT-5.5 fails on a subset of entries where other models recover signal, and entries where no evaluated model achieves non-zero F1 are dominated by structurally complex Go and Maven repositories.

The underlying failure mode is \textbf{\emph{signal dilution}}. In large repositories, vulnerability-relevant patterns such as unsafe calls, missing validation, or attacker-controlled data flow may appear in many benign contexts. The agent must therefore identify not only a suspicious pattern, but the specific instance that participates in the vulnerable implementation. This requires reading and integrating surrounding context across a scale that exceeds the effective working memory of current terminal agents. As a result, models often find plausible candidate files but fail to distinguish the security-critical implementation from syntactically similar benign code.

Distributed vulnerabilities create a related failure mode. When the ground truth spans multiple files, models frequently identify an initial vulnerable foothold but miss supporting files along the call path, configuration boundary, or validation chain. This explains why performance drops sharply on entries with many ground-truth files: the challenge is not merely finding one relevant file, but recovering the complete implementation slice that constitutes the vulnerability.

\paragraph{Does More Tool Use Help?}
\label{subsec:turn_budget}

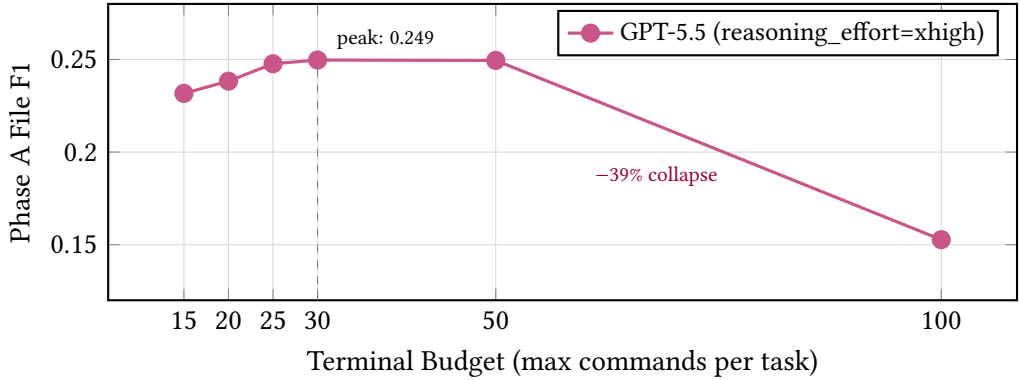
\begin{figure}[t]
\centering
\begin{tikzpicture}
\begin{axis}[
  width=0.85\linewidth,
  height=5.5cm,
  xlabel={Terminal Budget (max commands per task)},
  ylabel={Phase A File F1},
  xtick={15,20,25,30,50,100},
  xticklabels={15,20,25,30,50,100},
  ymin=0.12, ymax=0.28,
  grid=major,
  grid style={gray!30},
  mark size=3pt,
  thick,
  legend style={at={(0.98,0.98)}, anchor=north east, font=\small},
  every axis plot/.append style={line width=1.2pt},
]
\addplot[color=gpt, mark=*, mark options={fill=gpt}]
  coordinates {(15,0.2317) (20,0.2383) (25,0.2477) (30,0.2497) (50,0.2495) (100,0.1527)};
\addlegendentry{GPT-5.5 (reasoning\_effort=xhigh)}

\draw[dashed, gray, thin] (axis cs:30,0.12) -- (axis cs:30,0.2497);
\node[font=\scriptsize, anchor=south west] at (axis cs:31,0.250) {peak: 0.249};
\node[font=\scriptsize, anchor=north west, purple!80!black] at (axis cs:60,0.198) {$-$39\% collapse};
\end{axis}
\end{tikzpicture}
\caption{GPT-5.5 File F1 as a function of terminal command budget. Performance saturates around 30 commands and degrades at 100 commands, suggesting that localization is limited by evidence prioritization rather than command count alone.}
\label{fig:turn_budget}
\end{figure}

A natural hypothesis is that these failures arise because agents simply need more terminal commands. Turn-budget experiments do not support this explanation. As shown in Figure~\ref{fig:turn_budget}, increasing the budget from 15 to 30 commands yields only a modest improvement for GPT-5.5, after which performance saturates. At a budget of 100 commands, performance drops sharply as the model over-explores and loses confidence in its candidate set.

This suggests that the limiting factor is not the number of available commands, but the model's ability to decide \textbf{\emph{where}} to look and when to stop. More interaction can even be harmful when the agent lacks a stable evidence-ranking strategy: additional grep hits, directory listings, and file reads expand the candidate set faster than the model can resolve it. Future progress will therefore require better evidence memory, candidate ranking, and multi-file reasoning rather than simply longer tool budgets.

\paragraph{Operational Efficiency}
\label{subsec:cost_latency}

\begin{figure}[h!]
    \centering
    \includegraphics[width=\linewidth]{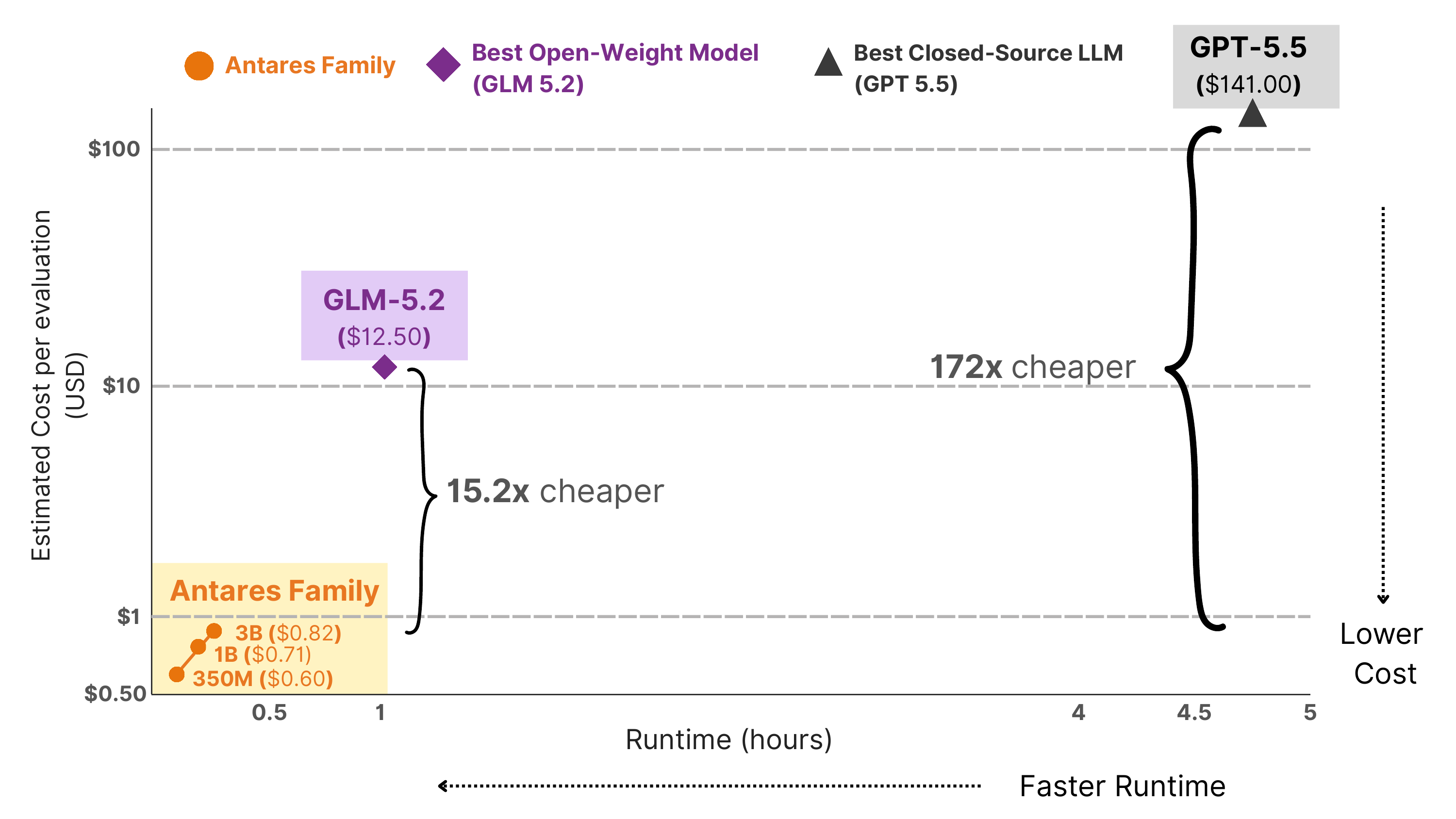}
    \caption{Runtime and estimated inference cost for evaluating the \vlb{}. All evaluations were performed using 16 parallel workers. Antares models were evaluated on a single H100 GPU, with costs estimated from publicly available H100 hourly rental prices. GLM-5.2 was evaluated through OpenRouter, while GPT-5.5 was evaluated through the OpenAI API. Reported runtime and cost reflect our evaluation setup and may vary depending on API pricing, deployment configuration, provider infrastructure, and rate limits.}
    \label{fig:runtime_cost_graph}
\end{figure}

Although these limitations remain, Antares completely changes the operational cost profile of repository-scale localization. Antares-3B completes the full 500-task evaluation sweep in approximately 15 minutes on a single H100 GPU with 16 parallel workers, corresponding to an estimated inference cost of less than \$1 per evaluation sweep using commodity H100 rental rates (Figure~\ref{fig:runtime_cost_graph}). In comparison, the strongest open-weight baseline, GLM-5.2, requires approximately 50 minutes and \$12.50 through OpenRouter, while GPT-5.5 requires approximately 5 hours and \$141 through the OpenAI API.

This gap matters because vulnerability localization is not usually a one-off query. Practical deployment requires repeated scans across repositories, branches, dependency updates, and CI/CD events. When deployed with a local inference endpoint, Antares enables repeated repository-scale evaluation without sending proprietary source code to third-party APIs, making the system suitable for closed-network security workflows where cost, latency, and source-code privacy are deployment constraints rather than secondary considerations.

%% file: content/10-safety.tex
\section{Safety and Responsible Disclosure}

\paragraph{Model release}
We release the Antares-350M and Antares-1B models on Hugging Face; Antares-3B, our most competitive variant, is retained for internal use and is not part of this release. Because Antares is a dual-use artifact---trained specifically to localize exploitable vulnerabilities in arbitrary repositories---we release the models with explicit acceptable-use restrictions. The models are intended for defensive security research, vulnerability assessment, remediation, and evaluation. We prohibit use for offensive cyber operations, unauthorized vulnerability discovery or exploitation, attacker enablement, credential theft, malware development, or any activity intended to compromise systems without authorization.

\paragraph{CLI release}
The Antares CLI is bundled with the Antares-1B Hugging Face repository rather than released independently. Because it operates over potentially
untrusted repositories and may connect to a user-configured inference endpoint,
the deployment is designed around repository containment and explicit data
boundaries. Before inference, eligible files are copied into a temporary read-only snapshot. Symlinks that resolve outside the repository are discarded, and the agent is restricted to parsed, read-only inspection commands. It cannot modify the repository, access the network, traverse beyond the snapshot, or inspect sensitive credential locations. Repository profiling used for automatic CWE selection is performed locally. Repository contents may themselves contain instructions intended to manipulate
the model. To reduce this risk, content returned through model-requested inspection is scanned before being added to the transcript, and detected
prompt-injection patterns are quarantined. Query and sweep operations may send the task instructions, repository paths, initial file inventory, and source content selected during inspection to the configured inference endpoint. Private local traces may also retain prompts, model responses, source excerpts,
commands, paths, and Git metadata until deletion.

%% file: content/11-conclusion.tex
\section{Conclusion}

We introduced Antares, a family of compact language models for agentic vulnerability localization. By combining cybersecurity reasoning, terminal exploration trajectories, and reinforcement learning from verifiable file-level rewards, Antares learns to localize vulnerable implementations directly from a repository and a CWE description, without relying on static-analysis candidates or frozen frontier-model scaffolds.

Our results show that targeted post-training can compensate for substantial differences in model scale. On \vlb{}, Antares-3B approaches GPT-5.5 while outperforming substantially larger open-weight models, and GRPO induces more stable, scale-dependent search strategies across the Antares family. At the same time, performance remains limited on large repositories, distributed vulnerabilities, and vulnerability classes defined by diffuse data flow, highlighting the need for stronger long-horizon repository reasoning.

Overall, Antares demonstrates that compact, locally deployable models can perform meaningful agentic security analysis when trained directly on the interaction pattern required by the task. To support continued research and responsible evaluation, we publicly release Antares-350M and Antares-1B through Hugging Face and bundle the inference CLI with the Antares-1B model repository.

%% file: content/acknowledgements.tex
\section*{Acknowledgements}

We thank Xuhong He, Karen Kui, Abhinav Chinta, Hadas Birin, Howard Lin, Huaibo Zhao, and Yasukazu Hirata for their support, guidance, and feedback throughout this work.

\noindent We also thank the S\&TO team, including Theo Morales, Aaron Carter, Thomas Bartlett, Omar Santos, and Anthony Grieco, for internally testing the model and providing feedback that helped inform the release process.

\noindent We are grateful to Jen Yokoyama, Marc Jones, and Elena Garcia from the legal team for their careful review and guidance.

\noindent We also thank Elizabeth Adkison and Emile Antone from the marketing team, as well as Susan O'Brien, Blake Thompson Heuer, Carro Halpin, and Nicole Greggs from the PR team, for their support with the release process and broader launch coordination.

%% file: content/appendix.tex

\section{Model and Evaluation Details}
\label{app:model_eval_details}

This section provides the implementation and behavioral details needed to
interpret the main evaluation. We first document the prompt and tool interface
used to ensure a controlled comparison across models. We then compare complete
Antares-3B and GPT-5.5 trajectories on the same repository to illustrate how
their search strategies differ under this interface. Finally, we compare the
agentic approach with conventional static-analysis tools to determine whether
interactive repository exploration contributes capabilities beyond rule-based
scanning.


\subsection{Evaluation Prompt and Tool Interface}
\label{app:evaluation_prompt}

A controlled agent interface is necessary to separate model capability from
differences in repository access or orchestration. Every evaluated model
therefore receives the same task instructions, repository state, terminal-call
budget, and submission protocol. Only the serialization of tool calls is
adapted to the model's native function-calling format.

The following system prompt is provided to every model during Vulnerability
Localization Benchmark evaluation.

\begin{promptbox}
You are a security vulnerability localization agent.
You have read-only terminal access to a codebase.

Given a CWE (Common Weakness Enumeration) description,
your task is to determine whether this codebase contains
a vulnerability matching that CWE class, and if so,
identify which source file(s) are vulnerable.

You can explore the codebase using the `terminal` tool.
It supports read-only commands such as ls, find, cat,
head, tail, grep, rg, and tree. You have up to 15
terminal calls.

When you are done exploring:
- If you found vulnerable files, call
  `submit_vulnerable_files` with a ranked list of exact
  repository-relative file paths.
- If you believe the codebase does not contain the
  described vulnerability, call
  `submit_no_vulnerability_found`.

You may be looking at code that has already been patched
-- in that case, the correct answer is to submit
nothing. Do not guess or hallucinate files. Only submit
files you have evidence for.

NOTE: Submitted paths must be exact file paths (e.g.
src/utils.js), never globs or wildcards.
\end{promptbox}

The user message supplies the CWE identifier and category description for the
current task together with the repository mount path. Each model receives at
most 15 terminal calls and must terminate through one of the two structured
submission tools.

\subsubsection{Tool Definitions}
\label{app:tool_definitions}

Three tools are provided to each model through JSON function schemas appended
to the system message.

\begin{toolbox}[title={Tool Definitions}]
{
  "type": "function",
  "function": {
    "name": "terminal",
    "description": "Execute a read-only terminal command in the repository. Supports standard file navigation, search, and inspection utilities. Read-only access only. Output is truncated to max_chars.",
    "parameters": {
      "type": "object",
      "properties": {
        "command": {
          "type": "string",
          "description": "The shell command to run"
        },
        "max_chars": {
          "type": "integer",
          "description": "Maximum number of output characters before truncation (default: 2000)",
          "default": 2000
        }
      },
      "required": ["command"]
    }
  }
}

{
  "type": "function",
  "function": {
    "name": "submit_vulnerable_files",
    "description": "Submit your answer: a ranked list of file paths you believe contain the vulnerability. Paths relative to repository root.",
    "parameters": {
      "type": "object",
      "properties": {
        "ranked_files": {
          "type": "array",
          "items": {
            "type": "string"
          },
          "description": "Ordered list of file paths"
        }
      },
      "required": ["ranked_files"]
    }
  }
}

{
  "type": "function",
  "function": {
    "name": "submit_no_vulnerability_found",
    "description": "Declare that no vulnerability matching the CWE description was found in this codebase.",
    "parameters": {
      "type": "object",
      "properties": {},
      "required": []
    }
  }
}
\end{toolbox}

\begin{observationbox}[title={Evaluation Control}]
All models operate with the same information, repository permissions, command
budget, and submission requirements. Differences in localization performance
therefore reflect the model's repository-search and verification policy rather
than differences in available tools or contextual information.
\end{observationbox}


\subsection{How Do Antares and GPT-5.5 Search the Same Repository?}
\label{app:trace_comparison}

\paragraph{Motivation.}
Aggregate File F1 indicates whether a model submits the correct files, but it
does not reveal how the model allocates its terminal budget, follows repository
structure, or distinguishes core implementations from nearby wrappers. We
therefore compare complete Antares-3B and GPT-5.5 trajectories on the same
benchmark task.

\paragraph{Example selection.}
We examine \texttt{trailofbits/fickling}
(GHSA-p523-jq9w-64x9; CVE-2026-22607), a Python pickle-safety
library containing vulnerabilities associated with CWE-184 and CWE-502. The
ground-truth vulnerable files are \texttt{fickling/fickle.py}, which implements
the core pickle interpreter, and \texttt{fickling/analysis.py}, which implements
the corresponding safety-analysis rules.

We select this task because both models locate relevant portions of the
repository but produce different final predictions. It therefore provides a
useful view of how search decisions affect file-level localization. This
example is intended as a qualitative illustration rather than evidence of an
aggregate model advantage. No repositories from the evaluation set appear in
the Antares training corpus.

\begin{table}[H]
\centering
\caption{Evaluation outcome on task \texttt{lemCG1XY}. Antares-3B identifies
both ground-truth files, whereas GPT-5.5 identifies one ground-truth file and
one adjacent delegation wrapper.}
\label{tab:trace_comparison_outcome}
\renewcommand{\arraystretch}{1.18}
\setlength{\tabcolsep}{8pt}
\begin{tabular}{lcccc}
\toprule
\textbf{Model}
& \textbf{Submitted Files}
& \textbf{Prec.}
& \textbf{Rec.}
& \textbf{F1} \\
\midrule
\rowcolor{antares}
\textbf{Antares-3B}
& \texttt{fickle.py, analysis.py}
& \textbf{1.0}
& \textbf{1.0}
& \textbf{1.0} \\
GPT-5.5 (xhigh)
& \texttt{analysis.py, loader.py}
& 0.5
& 0.5
& 0.5 \\
\bottomrule
\end{tabular}
\end{table}

\paragraph{Antares-3B trajectory.}
Antares-3B uses 14 terminal calls. The full trace below includes its explicit
reasoning blocks, tool calls, and terminal observations.

\input{content/traces/antares-3b_trace}

\paragraph{GPT-5.5 trajectory.}
GPT-5.5 uses the complete 15-call terminal budget. Because the API does not
expose its internal chain of thought, the trace contains tool calls and
observations but not the model's private reasoning process.

\input{content/traces/gpt-5.5_trace}

\paragraph{Comparison.}
Both models identify \texttt{fickling/analysis.py} and encounter references to
the underlying interpreter implementation. Antares follows the deprecated
\texttt{fickling/pickle.py} compatibility module to
\texttt{fickling/fickle.py}, reads the interpreter implementation, and submits
both ground-truth files. GPT-5.5 instead spends more of its budget examining
the public loading and hook interfaces. It submits
\texttt{fickling/loader.py}, which delegates to the safety-analysis pipeline,
but does not submit the underlying interpreter file.

\begin{observationbox}[title={Trace-Level Takeaway}]
On this example, the decisive difference is not whether the models discover
security-relevant files, but whether they trace wrapper and delegation layers
to the implementation that contains the vulnerable behavior. Antares completes
this implementation-level verification before submitting, whereas GPT-5.5
exhausts its terminal budget while investigating adjacent interface code. This
trace illustrates the search--verify--refine behavior analyzed quantitatively
in \autoref{subsec:behavioral_strategy}; it should not be interpreted as a
general model ranking from a single example.
\end{observationbox}


\subsection{How Does Agentic Localization Compare with Static Analysis?}
\label{app:sast_comparison}

\paragraph{Motivation.}
The main results compare Antares with language-model agents, but static analysis
remains the conventional approach to automated vulnerability detection. We
therefore ask whether interactive, CWE-conditioned repository exploration
provides useful localization signal beyond predefined rules and data-flow
queries.

\paragraph{Setup.}
We evaluate four static-analysis configurations on all 500 \vlb{} tasks. Broad Semgrep uses its default
\texttt{auto} configuration. A second, CWE-targeted Semgrep configuration maps
each benchmark task to an available CWE-specific rule pack. CodeQL constructs
a database for each repository and runs the
\texttt{security-extended} query suite for its primary language. Horusec invokes
its bundled language-specific analyzers. Findings from every tool are converted
to repository-relative file paths and evaluated using the same file-level
precision, recall, and F1 metrics as the agentic models.

\begin{table}[H]
\centering
\caption{Static-analysis performance on the 500-task Vulnerability
Localization Benchmark. Broad Semgrep provides the strongest static baseline,
but all configurations remain below the Antares family.}
\label{tab:sast_comparison}
\renewcommand{\arraystretch}{1.18}
\setlength{\tabcolsep}{9pt}
\begin{tabular}{lccc}
\toprule
\textbf{Configuration}
& \textbf{File F1}
& \textbf{Precision}
& \textbf{Recall} \\
\midrule
Semgrep (\texttt{auto}) & 0.086 & 0.091 & 0.155 \\
Semgrep (CWE-targeted) & 0.052 & 0.057 & 0.071 \\
CodeQL & 0.023 & 0.025 & 0.030 \\
Horusec & 0.020 & 0.021 & 0.038 \\
\bottomrule
\end{tabular}
\end{table}

\paragraph{Results.}
Broad Semgrep is the strongest static-analysis baseline, reaching 0.086 File
F1 with 0.155 recall. CWE-targeted Semgrep reaches 0.052 File F1 because
available rule packs cover only a subset of the 147 CWE categories and ecosystem
combinations represented in the benchmark. Restricting the rule set therefore
removes generic matches without producing a corresponding precision gain.

CodeQL and Horusec reach 0.023 and 0.020 File F1, respectively. CodeQL also
abstains on more than 92\% of entries, primarily because database construction
fails when extracted repository snapshots do not contain the required build
environment.

\begin{observationbox}[title={Static-Analysis Takeaway}]
Static-analysis tools recover vulnerabilities that match supported local
patterns, but their coverage depends on available rules, language front ends,
and successful project construction. Antares instead conditions its search on
the supplied CWE description and the structure of the current repository. Its
advantage is therefore largest when localization requires adapting the search
strategy across ecosystems or combining evidence from multiple files.
\end{observationbox}


\section{Additional Evaluation Benchmarks}
\label{app:additional_benchmarks}

The main evaluation measures the capability directly optimized during Antares
training: repository-scale vulnerability localization. This leaves two
questions unanswered. First, does the learned repository-navigation policy
transfer to non-security code-localization tasks? Second, does training on
extended terminal trajectories improve sequential tool use beyond the small
set of tools seen during training? We evaluate these questions using
issue-driven code localization and structured function calling.


\subsection{Does Vulnerability-Localization Training Transfer to General Code Search?}
\label{app:codescout_transfer}

\paragraph{Motivation.}
Vulnerability localization and issue-driven code localization provide different
task descriptions, but both require an agent to search an unfamiliar repository,
identify candidate files, and verify which implementations are relevant. We
test whether the search policy learned from security tasks transfers to this
broader code-localization setting.

\paragraph{Setup.}
We evaluate Antares on the CodeScout evaluation protocol
\cite{sutawika2026codescout} over SWE-Bench Verified
\cite{openai2024swebenchverified} and SWE-Bench Lite
\cite{jimenez2024swebench}. Given a GitHub issue description and the
pre-resolution repository state, the model must identify the files that require
modification.

Antares receives no additional fine-tuning and has no training exposure to
SWE-Bench repositories, GitHub issue descriptions, or issue-resolution
localization examples. Supervised fine-tuning includes general code-navigation
trajectories, but no bug-report-driven file identification of the type evaluated
here.

All Antares models use the OpenHands-Bash harness, and each result is averaged
over five independent runs. Published baseline results are included as reported
by CodeScout and do not provide evaluation variance.

\begin{table}[H]
\centering
\small
\caption{File-level localization performance on SWE-Bench Verified
(500 instances). Antares transfers to issue-driven localization without
SWE-Bench-specific training.}
\label{tab:codescout_verified}
\renewcommand{\arraystretch}{1.18}
\setlength{\tabcolsep}{5pt}
\begin{tabular}{llrccc}
\toprule
\textbf{Harness}
& \textbf{LLM}
& \textbf{Params}
& \textbf{File F1 (\%)}
& \textbf{Prec. (\%)}
& \textbf{Rec. (\%)} \\
\midrule
RepoNavigator
& Claude-Sonnet-4.5$^\dagger$
& --
& 79.94
& --
& -- \\

OpenHands-Bash
& CodeScout-14B (GRPO)$^\dagger$
& 14B
& 68.57
& 71.00
& 68.69 \\

OpenHands-Bash
& CodeScout-4B (GRPO)$^\dagger$
& 4B
& 68.52
& 71.53
& 67.74 \\

RepoNavigator
& Qwen2.5-32B (GRPO)$^\dagger$
& 32B
& 67.75
& 70.76
& 67.29 \\

\addlinespace[2pt]
\rowcolor{antares}
\textbf{OpenHands-Bash}
& \textbf{Antares-3B}
& \textbf{3B}
& \textbf{66.54 $\pm$ 0.22}
& \textbf{66.82 $\pm$ 0.25}
& \textbf{66.27 $\pm$ 0.20} \\
\addlinespace[2pt]

\rowcolor{antares}
\textbf{OpenHands-Bash}
& \textbf{Antares-1B}
& \textbf{1B}
& \textbf{64.24 $\pm$ 0.62}
& \textbf{64.51 $\pm$ 0.58}
& \textbf{63.97 $\pm$ 0.67} \\
\addlinespace[2pt]

OpenHands-Bash
& Qwen3-32B (Thinking)$^\dagger$
& 32B
& 62.91
& 59.87
& 73.63 \\

RepoNavigator
& Qwen2.5-14B (GRPO)$^\dagger$
& 14B
& 58.90
& 58.97
& 61.60 \\

RepoSearcher
& GPT-5-Chat$^\dagger$
& --
& 58.88
& 61.87
& 58.17 \\

OrcaLoca
& Qwen2.5-32B$^\dagger$
& 32B
& 58.11
& 59.51
& 59.57 \\

OpenHands-Bash
& CodeScout-1.7B (GRPO)$^\dagger$
& 1.7B
& 55.46
& 58.40
& 54.27 \\

RepoNavigator
& Qwen2.5-7B (GRPO)$^\dagger$
& 7B
& 51.63
& 53.83
& 50.62 \\

OpenHands-Bash
& Qwen3-4B-Instruct$^\dagger$
& 4B
& 49.73
& 49.69
& 53.34 \\

\addlinespace[2pt]
\rowcolor{antares}
\textbf{OpenHands-Bash}
& \textbf{Antares-350M}
& \textbf{350M}
& \textbf{49.65 $\pm$ 1.22}
& \textbf{49.88 $\pm$ 1.15}
& \textbf{49.42 $\pm$ 1.30} \\
\addlinespace[2pt]

OpenHands-Bash
& CodeScout-1.7B-RFT$^\dagger$
& 1.7B
& 46.60
& 48.60
& 45.82 \\

LocAgent
& Qwen2.5-32B$^\dagger$
& 32B
& 44.18
& 34.18
& 79.39 \\

OpenHands-Bash
& Qwen3-14B$^\dagger$
& 14B
& 43.13
& 36.49
& 71.20 \\

Agentless
& Qwen2.5-32B$^\dagger$
& 32B
& 35.38
& 25.60
& 78.93 \\

RepoSearcher
& Claude-3.7-Sonnet$^\dagger$
& --
& 32.30
& 20.24
& 89.24 \\

RepoSearcher
& Qwen2.5-32B (RFT)$^\dagger$
& 32B
& 32.25
& 20.24
& 88.59 \\

CoSIL
& Qwen2.5-32B$^\dagger$
& 32B
& 30.77
& 19.34
& 83.50 \\

RepoSearcher
& Qwen2.5-7B (RFT)$^\dagger$
& 7B
& 30.09
& 18.80
& 83.11 \\

OpenHands-Bash
& Qwen3-1.7B$^\dagger$
& 1.7B
& 2.40
& 2.09
& 3.60 \\
\bottomrule
\multicolumn{6}{l}{
\footnotesize
$^\dagger$ Results reported by \cite{sutawika2026codescout};
evaluation variance was not disclosed.
}
\end{tabular}
\end{table}

\begin{table}[H]
\centering
\small
\caption{File-level localization performance on SWE-Bench Lite
(300 instances). The transfer pattern remains consistent across the smaller
evaluation split.}
\label{tab:codescout_lite}
\renewcommand{\arraystretch}{1.18}
\setlength{\tabcolsep}{5pt}
\begin{tabular}{llrccc}
\toprule
\textbf{Harness}
& \textbf{LLM}
& \textbf{Params}
& \textbf{File F1 (\%)}
& \textbf{Prec. (\%)}
& \textbf{Rec. (\%)} \\
\midrule
OpenHands-Bash
& CodeScout-14B (GRPO)$^\dagger$
& 14B
& 71.84
& 71.17
& 73.36 \\

OpenHands-Bash
& CodeScout-4B (GRPO)$^\dagger$
& 4B
& 67.03
& 66.61
& 67.88 \\

\addlinespace[2pt]
\rowcolor{antares}
\textbf{OpenHands-Bash}
& \textbf{Antares-3B}
& \textbf{3B}
& \textbf{64.11 $\pm$ 0.59}
& \textbf{63.39 $\pm$ 0.61}
& \textbf{63.60 $\pm$ 0.58} \\
\addlinespace[2pt]

\rowcolor{antares}
\textbf{OpenHands-Bash}
& \textbf{Antares-1B}
& \textbf{1B}
& \textbf{61.74 $\pm$ 0.53}
& \textbf{61.20 $\pm$ 0.55}
& \textbf{61.53 $\pm$ 0.52} \\
\addlinespace[2pt]

OpenHands-Bash
& Qwen3-32B (Thinking)$^\dagger$
& 32B
& 58.98
& 54.26
& 71.53 \\

OpenHands-Bash
& CodeScout-1.7B (GRPO)$^\dagger$
& 1.7B
& 56.57
& 56.57
& 56.57 \\

OpenHands-Bash
& Gemma-4-31B
& 31B
& 48.36
& --
& -- \\

OpenHands-Bash
& Qwen3-4B-Instruct$^\dagger$
& 4B
& 47.41
& 43.70
& 55.47 \\

OpenHands-Bash
& CodeScout-1.7B-RFT$^\dagger$
& 1.7B
& 45.99
& 45.99
& 45.99 \\

\addlinespace[2pt]
\rowcolor{antares}
\textbf{OpenHands-Bash}
& \textbf{Antares-350M}
& \textbf{350M}
& \textbf{46.89 $\pm$ 1.17}
& \textbf{45.67 $\pm$ 1.20}
& \textbf{45.67 $\pm$ 1.15} \\
\addlinespace[2pt]

OpenHands-Bash
& GPT-OSS-20B
& 20B
& 44.28
& --
& -- \\

OpenHands-Bash
& Gemma-4-E2B
& 2B
& 39.71
& --
& -- \\

OpenHands-Bash
& Qwen3-14B$^\dagger$
& 14B
& 38.63
& 31.30
& 71.90 \\

OpenHands-Bash
& Gemma-4-E4B
& 4B
& 36.01
& --
& -- \\

OpenHands-Bash
& GPT-OSS-120B
& 120B
& 32.18
& --
& -- \\

LocAgent
& Claude-3.5-Sonnet$^\dagger$
& --
& 31.39
& 18.83
& 94.16 \\

OpenHands-Bash
& Qwen3-1.7B$^\dagger$
& 1.7B
& 2.16
& 1.96
& 2.92 \\

OpenHands-Bash
& GPT-5$^\dagger$
& --
& 1.09
& 1.09
& 1.09 \\

OpenHands-Bash
& Claude-Sonnet-4.5$^\dagger$
& --
& 0.36
& 0.36
& 0.36 \\
\bottomrule
\multicolumn{6}{l}{
\footnotesize
$^\dagger$ Results reported by \cite{sutawika2026codescout};
evaluation variance was not disclosed.
}
\end{tabular}
\end{table}

\paragraph{Results.}
On SWE-Bench Verified, Antares-3B reaches 66.54 File F1, within
2.03 points of CodeScout-14B and 1.98 points of CodeScout-4B, both
of which are directly optimized for SWE-Bench localization. Antares-1B reaches
64.24 and exceeds Qwen3-32B Thinking at 62.91. Antares-350M reaches
49.65, outperforming CodeScout-1.7B-RFT and approximately matching
Qwen3-4B-Instruct.

The same ordering largely holds on SWE-Bench Lite. Antares-3B reaches
64.11 File F1 and Antares-1B reaches 61.74, both exceeding
Qwen3-32B Thinking at 58.98. Antares-350M reaches 46.89,
remaining competitive with models several times larger.

\begin{observationbox}[title={Code-Localization Transfer}]
Antares transfers from CWE-conditioned vulnerability localization to
issue-driven file localization without task-specific adaptation. The transfer
suggests that reinforcement learning improves a reusable repository-navigation
policy---broad search, candidate verification, and iterative refinement---rather
than only learning security-specific lexical patterns.
\end{observationbox}


\subsection{Does Agentic Training Transfer to General Tool Calling?}
\label{app:bfcl_transfer}

\paragraph{Motivation.}
Antares is trained with only a small set of repository tools, but each rollout
requires the model to maintain state and select actions across as many as
15 turns. We ask whether this sequential interaction training improves
structured tool use when the model encounters unfamiliar function schemas.

\paragraph{Setup.}
We evaluate the Antares family on the Berkeley Function Calling Leaderboard v3
(BFCL-v3) \cite{pmlr-v267-patil25a}, which measures executable function
calling, live API abstract-syntax-tree accuracy, hallucination handling, and
multi-turn orchestration.

Antares receives no BFCL-style supervision and is trained with only three to
five fixed tools, whereas BFCL contains a much broader range of APIs and
function signatures. We report the overall score, the multi-turn orchestration
score, and Live-AST accuracy.

\begin{table}[H]
\centering
\caption{BFCL-v3 results. Antares remains close to its Granite base models on
the aggregate score while improving substantially on multi-turn orchestration.}
\label{tab:bfcl_results}
\renewcommand{\arraystretch}{1.18}
\setlength{\tabcolsep}{6pt}
\begin{tabular}{lrccc}
\toprule
\textbf{Model}
& \textbf{Params}
& \textbf{Overall}
& \textbf{Multi-Turn}
& \textbf{Live-AST} \\
\midrule
Qwen3.5-122B-A10B & 125B & 43.64 & 60.75 & 80.61 \\
Qwen3.5-35B-A3B & 36B & 41.81 & 56.25 & 79.42 \\
Qwen3.5-9B & 10B & 38.14 & 46.25 & 78.02 \\
GPT-OSS-120B & 120B & 32.16 & 45.38 & 67.21 \\
GPT-OSS-20B & 20B & 30.02 & 37.00 & 68.39 \\

\addlinespace[2pt]
\rowcolor{antares}
\textbf{Antares-3B}
& \textbf{3B}
& \textbf{28.64}
& \textbf{36.38}
& \textbf{65.51} \\
\addlinespace[2pt]

Granite-4.0-Micro & 3B & 27.94 & 20.38 & 54.63 \\
Llama-3.3-70B & 70B & 27.83 & 19.88 & 76.76 \\
Gemma-4-31B & 31B & 26.41 & 3.63 & 76.24 \\
GLM-4.7-Flash & 30B & 26.34 & 3.75 & 78.46 \\

\addlinespace[2pt]
\rowcolor{antares}
\textbf{Antares-1B}
& \textbf{1B}
& \textbf{26.26}
& \textbf{40.75}
& \textbf{61.95} \\
\addlinespace[2pt]

Gemma-4-26B-A4B & 26B & 25.42 & 1.25 & 68.02 \\
Granite-4.0-1B & 1B & 24.69 & 16.88 & 39.45 \\
Gemma-4-E2B & 2B & 24.68 & 9.38 & 74.61 \\
Qwen3.5-2B & 2B & 24.06 & 13.37 & 67.21 \\
Llama-Primus-Reasoning & 8B & 23.26 & 6.62 & 64.25 \\
Llama-3.2-3B & 3B & 20.38 & 3.88 & 58.11 \\

\addlinespace[2pt]
\rowcolor{antares}
\textbf{Antares-350M}
& \textbf{350M}
& \textbf{17.48}
& \textbf{24.63}
& \textbf{36.34} \\
\addlinespace[2pt]

Granite-4.0-350M & 350M & 17.29 & 2.50 & 33.53 \\
Foundation-Sec-8B & 8B & 10.00 & 0.00 & 0.00 \\
DeepHat-V1-7B & 7.6B & 9.99 & 0.00 & 0.00 \\
\bottomrule
\end{tabular}
\end{table}

\paragraph{Results.}
The overall BFCL scores of Antares remain close to those of the corresponding
Granite base models. The largest differences appear in multi-turn
orchestration. Antares-1B improves from 16.88 to 40.75,
Antares-3B improves from 20.38 to 36.38, and Antares-350M
improves from 2.50 to 24.63.

Antares-1B ranks fifth among the evaluated models on the multi-turn category
and exceeds GPT-OSS-20B despite containing twenty times fewer parameters.
Antares-350M exceeds every evaluated non-Antares model at 3B parameters or below in the reported multi-turn comparison.

Antares-3B additionally improves Live-AST accuracy from 54.63 for
Granite-4.0-Micro to 65.51. The smaller improvement in the aggregate score
indicates that the transfer is concentrated in sequential orchestration rather
than all forms of function-calling accuracy.

\begin{observationbox}[title={Tool-Use Transfer}]
Multi-turn trajectory training produces gains that extend beyond the repository
tools seen during Antares training. The improvements are concentrated in
maintaining coherent state and selecting tools across successive interactions,
which is the capability most directly exercised by the vulnerability-localization
agent loop.
\end{observationbox}


\section{Additional Analysis}
\label{app:additional_analysis}

This section tests the robustness and interpretation of the main findings. We
first examine whether localization difficulty varies across additional
benchmark dimensions. We then measure sensitivity to two evaluation choices:
the strategy encouraged by the system prompt and the orchestration provided by
the agent harness. Together, these analyses distinguish properties of the
learned model from properties introduced at inference time.


\subsection{Which Benchmark Dimensions Explain Localization Difficulty?}
\label{app:dimensional}

\paragraph{Motivation.}
The main paper shows that package ecosystem and repository scale strongly affect
localization performance, whereas CVSS severity does not. We provide two
additional views to determine whether this pattern is better explained by
programming-language structure or by vulnerability class.

\paragraph{Setup.}
We disaggregate File F1 by primary programming language and CWE category using
the same 500 tasks, model set, and evaluation protocol as the main
dimensional analysis in \autoref{subsec:dimensional_analysis}.

\subsubsection{Performance by Language}
\label{app:performance_by_language}

\begin{figure}[H]
\centering
\includegraphics[width=\linewidth]{
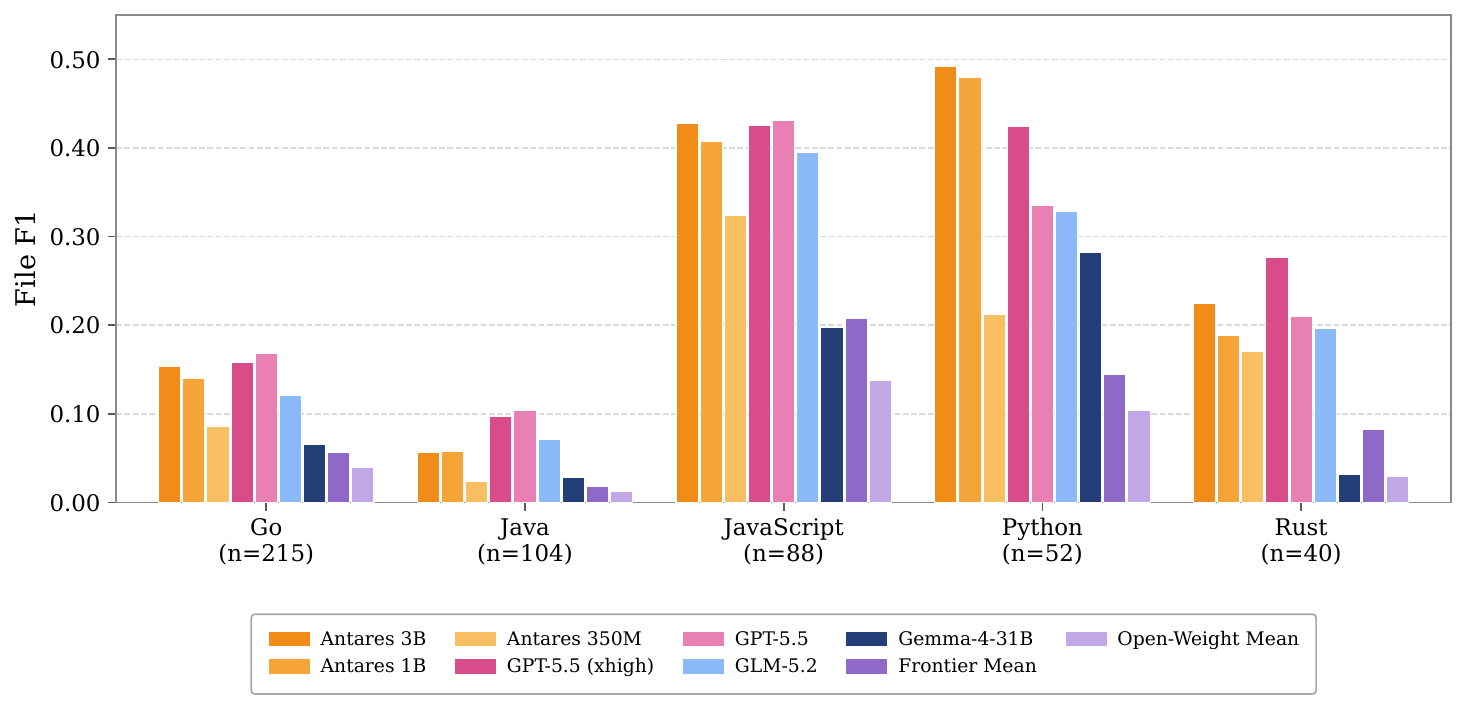
}
\caption{File F1 disaggregated by primary source language for the five most
frequent languages. Task counts: Go ($n=215$), Java ($n=104$),
JavaScript ($n=88$), Python ($n=52$), and Rust ($n=40$).}
\label{fig:perf_by_language}
\end{figure}

\paragraph{Results.}
Python and JavaScript yield substantially higher File F1 than Java for nearly
all evaluated models. Antares-3B reaches 0.492 on Python and 0.428 on
JavaScript, while GPT-5.5 retains an advantage on Rust and Java.

This pattern is consistent with differences in repository organization. Python
and JavaScript projects in the benchmark tend to expose relevant logic through
shallower directory structures, whereas Java projects frequently distribute
implementations across nested packages and framework layers.

\subsubsection{Performance by CWE Category}
\label{app:performance_by_cwe}

\begin{figure}[H]
\centering
\includegraphics[width=\linewidth]{
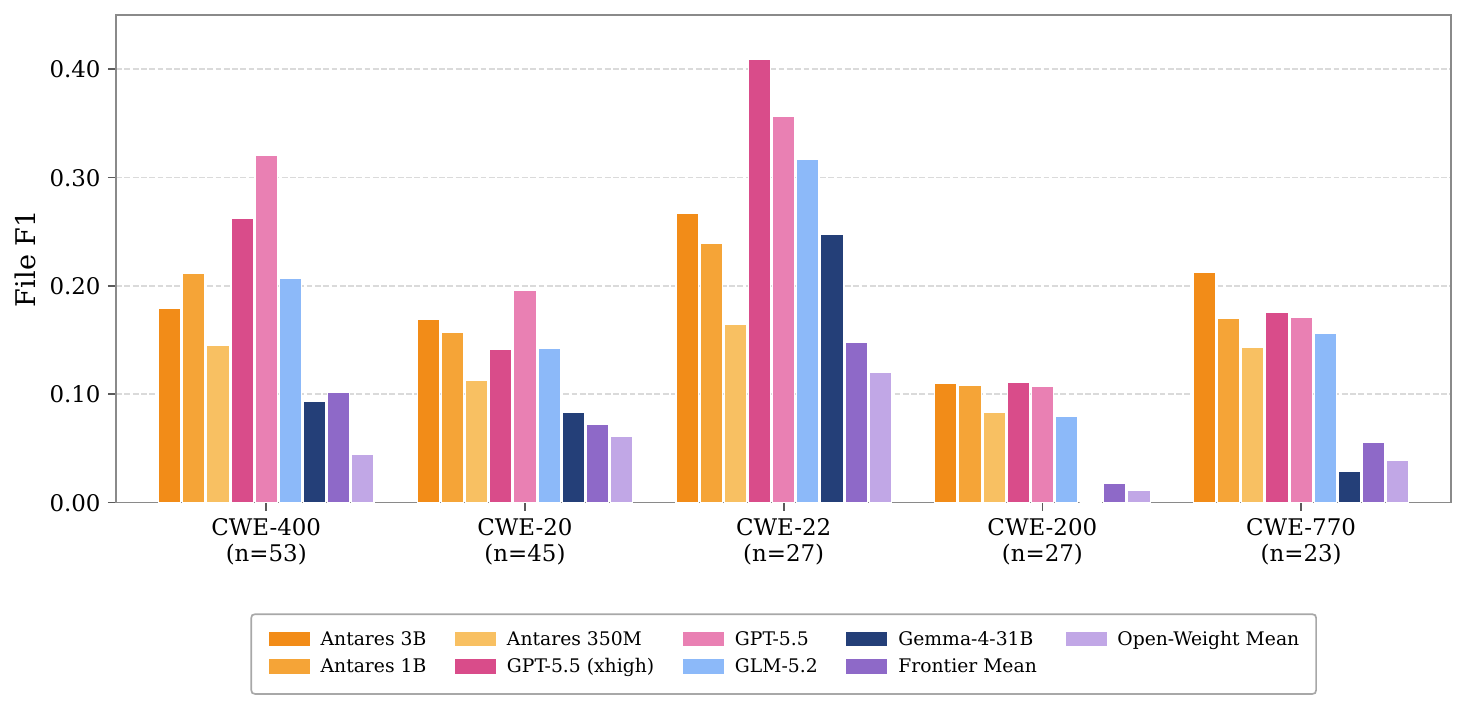
}
\caption{File F1 disaggregated by CWE category for the five most frequent
categories. Task counts: CWE-400 ($n=53$), CWE-20 ($n=45$),
CWE-22 ($n=27$), CWE-200 ($n=27$), and CWE-770 ($n=23$).}
\label{fig:perf_by_cwe}
\end{figure}

\paragraph{Results.}
Variation across CWE categories is considerably larger than variation across
CVSS severity bins. CWE-22 produces relatively strong localization performance
across models because path-processing logic often has recognizable lexical and
structural signatures. CWE-200 remains difficult because information exposure
can arise from diffuse data flow rather than a single distinctive
implementation pattern.

Antares-3B performs strongly on CWE-20 and CWE-770, while GPT-5.5 variants
retain an advantage on CWE-400 and CWE-22. These differences suggest that
models vary not only in aggregate localization quality, but also in the types
of structural evidence their search policies exploit effectively.

\begin{observationbox}[title={Dimensional Takeaway}]
Localization difficulty is driven more strongly by implementation structure
than by vulnerability impact. Languages with predictable repository layouts
and CWE categories with localized signatures are easier to search within a
fixed terminal budget. Diffuse vulnerabilities and deeply layered codebases
remain difficult across model families.
\end{observationbox}


\subsection{How Sensitive Is Antares to the System Prompt?}
\label{app:behavior_transfer}

\paragraph{Motivation.}
The behavioral analysis in \autoref{subsec:behavioral_strategy} shows that
Antares-3B follows a search-dominant strategy, whereas GPT-5.5 performs more
structural exploration before searching. We test whether this difference is a
fixed property of the learned policy or whether an explore-first strategy can
be elicited through instructions alone.

\paragraph{Setup.}
We modify only the Antares-3B system prompt by adding the following three-phase
strategy:

\begin{enumerate}
    \item \textbf{Explore first} (3--4 calls): map the repository structure
    before searching and form an initial model of the codebase organization.

    \item \textbf{Targeted search} (4--6 calls): use the structural overview
    to search for vulnerability-relevant patterns in likely directories.

    \item \textbf{Verify and read} (3--5 calls): inspect candidate files and
    confirm that the vulnerability is implemented in the submitted paths.
\end{enumerate}

The model checkpoint, inference parameters, tool definitions, sandbox, scoring
function, and evaluation entries remain unchanged. No retraining or gradient
updates are performed.

\begin{table}[H]
\centering
\caption{Command-distribution comparison across behavioral profiles. The
prompt-modified model moves toward an explore-first strategy while retaining
most of the baseline model's search efficiency.}
\label{tab:behavior_transfer}
\renewcommand{\arraystretch}{1.18}
\setlength{\tabcolsep}{8pt}
\begin{tabular}{lccc}
\toprule
\textbf{Metric}
& \textbf{GPT-5.5 (xhigh)}
& \textbf{Explore-First}
& \textbf{Baseline 3B} \\
\midrule
List/explore commands
& 20.2\%
& 17.3\%
& 10.2\% \\

Grep/search commands
& 34.2\%
& 46.2\%
& 52.3\% \\

Opening
\texttt{ls}$\rightarrow$\texttt{ls}$\rightarrow$$\ast$
& 425/500
& 93/500
& 23/500 \\

Mean commands per task
& 15.8
& 14.6
& 13.9 \\
\bottomrule
\end{tabular}
\end{table}

\paragraph{Results.}
The explore-first prompt raises Antares-3B File F1 from 0.223 to
0.2313, slightly above the 0.2292 score obtained by GPT-5.5
(xhigh). Structural exploration increases from 10.22\% to 17.3\%, while
search commands decrease from 52.31\% to 46.2\%. The mean number of
commands rises only modestly, from 13.9 to 14.6.

The modified model therefore partially adopts GPT-5.5's structural exploration
profile without reproducing its larger command budget. The resulting behavior
combines broader initial orientation with the search-heavy strategy dominant in
the baseline Antares policy.

\begin{observationbox}[title={Prompt-Sensitivity Takeaway}]
Antares is sensitive to strategy-level instructions, but not merely at the
surface level: the prompt changes both its command distribution and its final
localization accuracy. The result indicates that explore-first behavior remains
available within the learned policy even though it is not dominant under the
baseline prompt.
\end{observationbox}

\paragraph{Possible explanation.}
One hypothesis is that the supervised fine-tuning initialization assigns more
probability to grep-dominant trajectories, causing GRPO to refine this behavior
rather than discover a distinct exploration-first mode. The present experiment
does not isolate the source of the behavior, but it motivates future training
with more diverse repository-navigation trajectories and rollout
initializations.


\subsection{How Sensitive Are Results to the Agent Harness?}
\label{app:harness_sensitivity}

\paragraph{Motivation.}
A standardized harness is required for controlled model comparison, but it may
not represent the strongest configuration available for each model. We examine
how much performance changes when the Antares harness is optimized and when a
frontier model is allowed to operate through its native agent interface.

\paragraph{Baseline configuration.}
All main-paper results use the same single-loop agent harness, 15-call terminal
budget, Docker sandbox, task prompt, and file-submission protocol. The harness
parses each model's native tool-calling output, executes terminal commands in
the sandbox, and returns stdout as tool observations. This standardization
isolates model differences but deliberately excludes model-specific
orchestration.

\subsubsection{Antares Harness Optimization}
\label{app:fapo_harness}

We apply FAPO (Fully Automated Prompt Optimization)
\cite{kassianik2026fapofullyautomatedprompt} to the Antares-3B agent harness.
FAPO iteratively evaluates a multi-step LLM pipeline, diagnoses failure modes
from intermediate outputs, proposes scoped prompt or configuration edits, and
validates the resulting variants against a target score.

Applied to Antares-3B, FAPO identifies a four-phase strategy:
\emph{orient}, \emph{narrow}, \emph{confirm}, and \emph{submit}. The optimized
configuration increases the terminal budget from 15 to 25 calls and modifies
several inference and loop parameters, including a frequency penalty of 0.3,
a maximum of 4,096 tokens per turn, and temperature 0.3.

The optimized configuration improves File F1 from 0.223 to
\textbf{0.235}, a 5.4\% relative gain, without changing the model
weights or architecture.

\subsubsection{Native Frontier-Agent Evaluation}
\label{app:native_frontier_agent}

We evaluate Claude Opus 4.6 through Claude Code on the same Vulnerability Localization Benchmark tasks in May 2026. Each task runs in a separate Claude Code instance using native tool orchestration and subagent spawning. We impose no explicit token or interaction limit. Unlike the standardized harness, this configuration provides
full native tool orchestration, subagent spawning, and unconstrained interaction
budgets.

The native frontier-agent configuration reaches 0.284 File F1 at a total
cost of approximately \$1,658, or \$3.32 per task, and requires approximately
2.3 minutes per task. The reported cost sums the billed Claude Code usage across the completed evaluation. Antares-3B averages 1.96 seconds of model generation and approximately 26.5 seconds of orchestration and sandbox overhead per task. With 16 parallel workers, the full 500-task sweep completes in approximately 15 minutes, corresponding to an amortized throughput of under 2 seconds per task.

The native frontier configuration invokes subagents on 13.2\% of entries, a
capability absent from the standardized Antares harness. It operates at
approximately 1,660$\times$ greater cost and 57$\times$ longer
model-inference time. Twelve runs attempted to access evaluation metadata
outside the permitted repository context and were invalidated and rerun; these
metadata files are inaccessible through the standardized harness. The reported cost includes the 12 invalidated runs and their replacements.

\paragraph{Results.}
Harness optimization is a meaningful performance axis. FAPO raises
Antares-3B from 0.223 to 0.235 File F1 without changing the model
weights. The native frontier-agent configuration reaches 0.284, but requires
substantially greater inference time, cost, and orchestration flexibility.

Even under this larger inference budget, performance remains far below perfect
localization. These results indicate that substantially greater inference-time compute does not saturate the benchmark under the evaluated configuration.

\begin{observationbox}[title={Harness-Sensitivity Takeaway}]
The harness is a meaningful source of performance variation: orchestration
changes improve Antares without retraining and allow frontier agents to achieve
higher absolute accuracy. However, these gains do not eliminate the efficiency
difference, and the benchmark remains unsaturated even under an unconstrained,
high-cost frontier-agent configuration.
\end{observationbox}

%% file: content/traces/antares-3b_trace.tex
\begin{tcolorbox}[
    title=Antares-3B Execution Trace,
    colback=white,
    colframe=pink!50!black,
    fonttitle=\bfseries,
    breakable
]

\begin{lstlisting}[
    basicstyle=\ttfamily\small,
    numbers=none,
    breaklines=true,
    backgroundcolor=\color{orange!15}
]
<think>
Start by listing repository root to understand project
structure and likely languages/frameworks.
</think>
\end{lstlisting}

\textbf{Tool Call}
\begin{lstlisting}[
    basicstyle=\ttfamily\small,
    numbers=none,
    breaklines=true,
    backgroundcolor=\color{pink!30}
]
CALL: ls
\end{lstlisting}

\textbf{Observation}
\begin{lstlisting}[
    basicstyle=\ttfamily\small,
    numbers=none,
    breaklines=true,
    backgroundcolor=\color{black!3}
]
  -> CODEOWNERS
     DEVELOPMENT.md
     LICENSE
     Makefile
     README.md
     example
     fickling
     fickling_image.png
     pickle_scanning_benchmark
     pyproject.toml
     test
     uv.lock
     [14 tool-calls remaining]
\end{lstlisting}

\vspace{1em}
\begin{lstlisting}[
    basicstyle=\ttfamily\small,
    numbers=none,
    breaklines=true,
    backgroundcolor=\color{orange!15}
]
<think>
Find Python code and any deserialization or blacklist-based
validation logic relevant to pickle scanning/deserialization
vulnerabilities.
</think>
\end{lstlisting}

\textbf{Tool Call}
\begin{lstlisting}[
    basicstyle=\ttfamily\small,
    numbers=none,
    breaklines=true,
    backgroundcolor=\color{pink!30}
]
CALL: rg "pickle|load\(|loads|Blacklist|deny|disallow|unsafe|unsafe_pickle|scan" -n fickling test example | head -n 200
\end{lstlisting}

\textbf{Observation}
\begin{lstlisting}[
    basicstyle=\ttfamily\small,
    numbers=none,
    breaklines=true,
    backgroundcolor=\color{black!3}
]
  -> example/fault_injection.py:8:import pickle
     example/fault_injection.py:36:safe_model = Pickled.load(
       pickle.dumps(model))
     test/test_loads_api.py:2:import pickle
     test/test_loads_api.py:21:"Test the fickling.loads()..."
     ...
     [TRUNCATED -- 14852 total chars, showing first 2000]
     [13 tool-calls remaining]
\end{lstlisting}

\vspace{1em}
\begin{lstlisting}[
    basicstyle=\ttfamily\small,
    numbers=none,
    breaklines=true,
    backgroundcolor=\color{orange!15}
]
<think>
Locate the main implementation files for pickle
parsing/deserialization and safety checks.
</think>
\end{lstlisting}

\textbf{Tool Call}
\begin{lstlisting}[
    basicstyle=\ttfamily\small,
    numbers=none,
    breaklines=true,
    backgroundcolor=\color{pink!30}
]
CALL: find fickling -type f | sort
\end{lstlisting}

\textbf{Observation}
\begin{lstlisting}[
    basicstyle=\ttfamily\small,
    numbers=none,
    breaklines=true,
    backgroundcolor=\color{black!3}
]
  -> fickling/__init__.py
     fickling/__main__.py
     fickling/analysis.py
     fickling/cli.py
     fickling/context.py
     fickling/exception.py
     fickling/fickle.py
     fickling/hook.py
     fickling/import_hook.py
     fickling/loader.py
     fickling/ml.py
     fickling/pickle.py
     fickling/polyglot.py
     fickling/pytorch.py
     fickling/tracing.py
     [12 tool-calls remaining]
\end{lstlisting}

\vspace{1em}

\begin{lstlisting}[
    basicstyle=\ttfamily\small,
    numbers=none,
    breaklines=true,
    backgroundcolor=\color{orange!15}
]
<think>
Read core pickle implementation to see whether it fully
deserializes untrusted data or contains a denylist that
may be incomplete.
</think>
\end{lstlisting}

\textbf{Tool Call}
\begin{lstlisting}[
    basicstyle=\ttfamily\small,
    numbers=none,
    breaklines=true,
    backgroundcolor=\color{pink!30}
]
CALL: sed -n '1,240p' fickling/pickle.py
\end{lstlisting}

\textbf{Observation}
\begin{lstlisting}[
    basicstyle=\ttfamily\small,
    numbers=none,
    breaklines=true,
    backgroundcolor=\color{black!3}
]
  -> import warnings
     from fickling.fickle import *  # noqa: F403
     warnings.warn(
       "As of version 0.0.8, `fickling.pickle` has been
        renamed `fickling.fickle`. The `fickling.pickle`
        module will be removed in a future release!",
       DeprecationWarning, stacklevel=2,)
     [11 tool-calls remaining]
\end{lstlisting}

\vspace{1em}
\begin{lstlisting}[
    basicstyle=\ttfamily\small,
    numbers=none,
    breaklines=true,
    backgroundcolor=\color{orange!15}
]
<think>
Continue exploring public API module around deprecated/new
pickle API and validation behavior.
</think>
\end{lstlisting}

\textbf{Tool Call}
\begin{lstlisting}[
    basicstyle=\ttfamily\small,
    numbers=none,
    breaklines=true,
    backgroundcolor=\color{pink!30}
]
CALL: sed -n '240,520p' fickling/pickle.py
\end{lstlisting}

\textbf{Observation}
\begin{lstlisting}[
    basicstyle=\ttfamily\small,
    numbers=none,
    breaklines=true,
    backgroundcolor=\color{black!3}
]
  -> (no output)
     [10 tool-calls remaining]
\end{lstlisting}

\vspace{1em}

\begin{lstlisting}[
    basicstyle=\ttfamily\small,
    numbers=none,
    breaklines=true,
    backgroundcolor=\color{orange!15}
]
<think>
Check package exports to determine whether users call
actual deserialization APIs.
</think>
\end{lstlisting}

\textbf{Tool Call}
\begin{lstlisting}[
    basicstyle=\ttfamily\small,
    numbers=none,
    breaklines=true,
    backgroundcolor=\color{pink!30}
]
CALL: sed -n '1,220p' fickling/__init__.py
\end{lstlisting}

\textbf{Observation}
\begin{lstlisting}[
    basicstyle=\ttfamily\small,
    numbers=none,
    breaklines=true,
    backgroundcolor=\color{black!3}
]
  -> from .loader import load, loads #noqa
     from .context import check_safety #noqa
     from .hook import always_check_safety,
       activate_safe_ml_environment #noqa
     from .analysis import is_likely_safe # noqa
     __version__ = "0.1.5"
     [9 tool-calls remaining]
\end{lstlisting}

\vspace{1em}

\begin{lstlisting}[
    basicstyle=\ttfamily\small,
    numbers=none,
    breaklines=true,
    backgroundcolor=\color{orange!15}
]
<think>
Find all safety checks and deserialization entry points to
assess if untrusted data is deserialized with incomplete
protections.
</think>
\end{lstlisting}

\textbf{Tool Call}
\begin{lstlisting}[
    basicstyle=\ttfamily\small,
    numbers=none,
    breaklines=true,
    backgroundcolor=\color{pink!30}
]
CALL: rg "loads\(|load\(|check_safety|always_check_safety|is_likely_safe|analysis" fickling -n
\end{lstlisting}

\textbf{Observation}
\begin{lstlisting}[
    basicstyle=\ttfamily\small,
    numbers=none,
    breaklines=true,
    backgroundcolor=\color{black!3}
]
  -> hook.py:16: Drop-in replacement for pickle.Unpickler
       that uses fickling's safety analysis.
     hook.py:27: def load(self):
     hook.py:29: return loader.load(self._file, ...)
     cli.py:118: stacked_pickled =
       fickle.StackedPickle.load(file, ...)
     loader.py:24: return pickle.loads(
       pickled_data.dumps(), ...)
     ...
     [TRUNCATED -- 7631 total chars, showing first 2000]
     [8 tool-calls remaining]
\end{lstlisting}

\vspace{1em}

\begin{lstlisting}[
    basicstyle=\ttfamily\small,
    numbers=none,
    breaklines=true,
    backgroundcolor=\color{orange!15}
]
<think>
Inspect the safety analysis implementation for
denylist-based checks that might be incomplete.
</think>
\end{lstlisting}

\textbf{Tool Call}
\begin{lstlisting}[
    basicstyle=\ttfamily\small,
    numbers=none,
    breaklines=true,
    backgroundcolor=\color{pink!30}
]
CALL: sed -n '1,260p' fickling/analysis.py
\end{lstlisting}

\textbf{Observation}
\begin{lstlisting}[
    basicstyle=\ttfamily\small,
    numbers=none,
    breaklines=true,
    backgroundcolor=\color{black!3}
]
  -> from fickling.fickle import Interpreter, Pickled, Proto
     class AnalyzerMeta(type): ...
     class AnalysisContext: ...
     class Analyzer(metaclass=AnalyzerMeta): ...
     class Severity(Enum): ...
     class AnalysisResult: ...
     class Analysis(ABC): ...
     class DuplicateProtoAnalysis(Analysis): ...
     class MisplacedProtoAnalysis(Analysis): ...
     [TRUNCATED -- 11362 total chars, showing first 2000]
     [7 tool-calls remaining]
\end{lstlisting}

\vspace{1em}

\begin{lstlisting}[
    basicstyle=\ttfamily\small,
    numbers=none,
    breaklines=true,
    backgroundcolor=\color{orange!15}
]
<think>
Find specific analysis rules and any denylist/blocklist
patterns.
</think>
\end{lstlisting}

\textbf{Tool Call}
\begin{lstlisting}[
    basicstyle=\ttfamily\small,
    numbers=none,
    breaklines=true,
    backgroundcolor=\color{pink!30}
]
CALL: rg "Analysis\(|AnalysisResult|Severity\(|is_likely_safe|Analyzer" fickling/analysis.py -n
\end{lstlisting}

\textbf{Observation}
\begin{lstlisting}[
    basicstyle=\ttfamily\small,
    numbers=none,
    breaklines=true,
    backgroundcolor=\color{black!3}
]
  -> 13:class AnalyzerMeta(type):
     58:class Analyzer(metaclass=AnalyzerMeta):
     126:class Analysis(ABC):
     137:class DuplicateProtoAnalysis(Analysis):
     170:class MisplacedProtoAnalysis(Analysis):
     186:class InvalidOpcode(Analysis) [NonStandardImports]
     249:class UnsafeImportsML(Analysis):
     283:class BadCalls(Analysis):
     296:class OvertlyBadEvals(Analysis):
     ...
     [TRUNCATED -- 2731 total chars, showing first 2000]
     [6 tool-calls remaining]
\end{lstlisting}

\vspace{1em}

\begin{lstlisting}[
    basicstyle=\ttfamily\small,
    numbers=none,
    breaklines=true,
    backgroundcolor=\color{orange!15}
]
<think>
Read later analysis rules where disallowed
imports/functions/attributes are evaluated.
</think>
\end{lstlisting}

\textbf{Tool Call}
\begin{lstlisting}[
    basicstyle=\ttfamily\small,
    numbers=none,
    breaklines=true,
    backgroundcolor=\color{pink!30}
]
CALL: sed -n '300,620p' fickling/analysis.py
\end{lstlisting}

\textbf{Observation}
\begin{lstlisting}[
    basicstyle=\ttfamily\small,
    numbers=none,
    breaklines=true,
    backgroundcolor=\color{black!3}
]
  -> shortened.startswith("eval(")
     or shortened.startswith("exec(")
     or shortened.startswith("compile(")
     or shortened.startswith("open(")
     or shortened.startswith("_run_code(")
     or shortened.startswith("execWrapper(")
     ... yield AnalysisResult(Severity.OVERTLY_MALICIOUS,
       "Call to `{shortened}` is almost certainly evidence
        of a malicious pickle file", "OvertlyBadEval")
     class UnsafeImports(Analysis): ...
     class UnusedVariables(Analysis): ...
     [TRUNCATED -- 5006 total chars, showing first 2000]
     [5 tool-calls remaining]
\end{lstlisting}

\vspace{1em}

\begin{lstlisting}[
    basicstyle=\ttfamily\small,
    numbers=none,
    breaklines=true,
    backgroundcolor=\color{orange!15}
]
<think>
Continue through analysis rules for imports/attributes/calls
that are flagged as likely unsafe.
</think>
\end{lstlisting}

\textbf{Tool Call}
\begin{lstlisting}[
    basicstyle=\ttfamily\small,
    numbers=none,
    breaklines=true,
    backgroundcolor=\color{pink!30}
]
CALL: sed -n '620,1040p' fickling/analysis.py
\end{lstlisting}

\textbf{Observation}
\begin{lstlisting}[
    basicstyle=\ttfamily\small,
    numbers=none,
    breaklines=true,
    backgroundcolor=\color{black!3}
]
  -> (no output)
     [4 tool-calls remaining]
\end{lstlisting}

\vspace{1em}

\begin{lstlisting}[
    basicstyle=\ttfamily\small,
    numbers=none,
    breaklines=true,
    backgroundcolor=\color{orange!15}
]
<think>
Get complete analysis.py with line numbers to precisely
identify vulnerable rule implementations.
</think>
\end{lstlisting}

\textbf{Tool Call}
\begin{lstlisting}[
    basicstyle=\ttfamily\small,
    numbers=none,
    breaklines=true,
    backgroundcolor=\color{pink!30}
]
CALL: nl -ba fickling/analysis.py | sed -n '1,360p'
\end{lstlisting}

\textbf{Observation}
\begin{lstlisting}[
    basicstyle=\ttfamily\small,
    numbers=none,
    breaklines=true,
    backgroundcolor=\color{black!3}
]
  -> 1  from __future__ import annotations
     ...
     10 from fickling.fickle import Interpreter, Pickled, Proto
     ...
     296 class OvertlyBadEvals(Analysis):
     ... shortened.startswith("eval(") or "exec(" or
         "compile(" or "open(" or "_run_code(" or
         "execWrapper(" -- hardcoded fixed list
     ...
     [TRUNCATED -- 18342 total chars, showing first 2000]
     [3 tool-calls remaining]
\end{lstlisting}

\vspace{1em}

\begin{lstlisting}[
    basicstyle=\ttfamily\small,
    numbers=none,
    breaklines=true,
    backgroundcolor=\color{orange!15}
]
<think>
Locate all Analysis subclasses and their analyze() methods.
</think>
\end{lstlisting}

\textbf{Tool Call}
\begin{lstlisting}[
    basicstyle=\ttfamily\small,
    numbers=none,
    breaklines=true,
    backgroundcolor=\color{pink!30}
]
CALL: rg "class .*Analysis" fickling/analysis.py
\end{lstlisting}

\textbf{Observation}
\begin{lstlisting}[
    basicstyle=\ttfamily\small,
    numbers=none,
    breaklines=true,
    backgroundcolor=\color{black!3}
]
  -> class AnalysisContext:
     class AnalysisResult:
     class Analysis(ABC):
     class DuplicateProtoAnalysis(Analysis):
     class MisplacedProtoAnalysis(Analysis):
     class InvalidOpcode(Analysis):
     class NonStandardImports(Analysis):
     class UnsafeImportsML(Analysis):
     class BadCalls(Analysis):
     class OvertlyBadEvals(Analysis):
     class UnsafeImports(Analysis):
     class UnusedVariables(Analysis):
     class AnalysisResults:
     [2 tool-calls remaining]
\end{lstlisting}

\vspace{1em}

\begin{lstlisting}[
    basicstyle=\ttfamily\small,
    numbers=none,
    breaklines=true,
    backgroundcolor=\color{orange!15}
]
<think>
Read the interpreter implementation to see how it executes
opcodes and whether it safely handles untrusted pickle data.
</think>
\end{lstlisting}

\textbf{Tool Call}
\begin{lstlisting}[
    basicstyle=\ttfamily\small,
    numbers=none,
    breaklines=true,
    backgroundcolor=\color{pink!30}
]
CALL: sed -n '1,220p' fickling/fickle.py
\end{lstlisting}

\textbf{Observation}
\begin{lstlisting}[
    basicstyle=\ttfamily\small,
    numbers=none,
    breaklines=true,
    backgroundcolor=\color{black!3}
]
  -> from stdlib_list import in_stdlib
     from fickling.exception import WrongMethodError
     BUILTIN_MODULE_NAMES: frozenset[str] =
       frozenset(sys.builtin_module_names)
     def is_std_module(module_name: str) -> bool:
       return in_stdlib(module_name)
         or module_name in BUILTIN_MODULE_NAMES
     def extract_identifier_from_ast_node(node, ...): ...
     [TRUNCATED -- 7039 total chars, showing first 2000]
     [1 tool-calls remaining]
\end{lstlisting}

\vspace{1em}

\begin{lstlisting}[
    basicstyle=\ttfamily\small,
    numbers=none,
    breaklines=true,
    backgroundcolor=\color{orange!15}
]
<think>
fickling/fickle.py is the critical file: it implements the
unpickler/interpreter for untrusted pickle data. The safety
analysis in fickling/analysis.py is secondary and relies on
properties discovered by the interpreter/fickle. The combined
effect is a protection mechanism that depends on a list/set
of allowed/unsafe module names and analysis rules. An incomplete
or brittle list of disallowed imports or calls can let malicious
pickles pass the safety checks before or during deserialization.
</think>
\end{lstlisting}

\textbf{Final Tool Call}
\begin{lstlisting}[
    basicstyle=\ttfamily\small,
    numbers=none,
    breaklines=true,
    backgroundcolor=\color{pink!30}
]
CALL: submit_vulnerable_files
\end{lstlisting}

\textbf{Submission}
\begin{lstlisting}[
    basicstyle=\ttfamily\small,
    numbers=none,
    breaklines=true,
    backgroundcolor=\color{blue!8}
]
  -> ["fickling/fickle.py", "fickling/analysis.py"]
\end{lstlisting}

\vspace{1em}

\textbf{Result}
\begin{lstlisting}[
    basicstyle=\ttfamily\small,
    numbers=none,
    breaklines=true,
    backgroundcolor=\color{mushygreen!40}
]
Precision: 1.0
Recall:    1.0
File F1:   1.0
\end{lstlisting}
\end{tcolorbox}

%% file: content/traces/gpt-5.5_trace.tex
\begin{tcolorbox}[
    title=GPT-5.5 Execution Trace,
    colback=white,
    colframe=pink!50!black,
    fonttitle=\bfseries,
    breakable
]

\begin{lstlisting}[
    basicstyle=\ttfamily\small,
    numbers=none,
    breaklines=true,
    backgroundcolor=\color{orange!15}
]
[reasoning undisclosed]

\end{lstlisting}

\textbf{Tool Call}
\begin{lstlisting}[
    basicstyle=\ttfamily\small,
    numbers=none,
    breaklines=true,
    backgroundcolor=\color{pink!30}
]
CALL: cd /workspace/repo && find . -maxdepth 2 -type f \
      | sed 's#^./##' | sort | head -200
\end{lstlisting}

\textbf{Observation}
\begin{lstlisting}[
    basicstyle=\ttfamily\small,
    numbers=none,
    breaklines=true,
    backgroundcolor=\color{black!3}
]
  -> CODEOWNERS
     DEVELOPMENT.md
     LICENSE
     Makefile
     README.md
     example/context_manager.py
     example/fault_injection.py
     fickling/__init__.py
     fickling/analysis.py
     fickling/cli.py
     fickling/fickle.py
     fickling/hook.py
     fickling/loader.py
     fickling/pickle.py
     fickling/polyglot.py
     fickling/pytorch.py
     pickle_scanning_benchmark/benchmark.py
     test/test_pickle.py
     ...
     (53 files total)
\end{lstlisting}

\vspace{1em}

\begin{lstlisting}[
    basicstyle=\ttfamily\small,
    numbers=none,
    breaklines=true,
    backgroundcolor=\color{orange!15}
]
[reasoning undisclosed]

\end{lstlisting}

\textbf{Tool Call}
\begin{lstlisting}[
    basicstyle=\ttfamily\small,
    numbers=none,
    breaklines=true,
    backgroundcolor=\color{pink!30}
]
CALL: rg -n "(pickle\.|loads\(|load\(|Unpickler|torch\.load|yaml\.load|marshal\.loads|dill|cloudpickle|eval\(|exec\(|blacklist|deny|disallow|unsafe|safe|allowed|check_safety|check)" fickling test example | head -300
\end{lstlisting}

\textbf{Observation}
\begin{lstlisting}[
    basicstyle=\ttfamily\small,
    numbers=none,
    breaklines=true,
    backgroundcolor=\color{black!3}
]
  -> example/fault_injection.py:8:import pickle
     example/fault_injection.py:36:Pickled.load(pickle.dumps(...))
     fickling/hook.py:16:Drop-in replacement for pickle.Unpickler
       that uses fickling's safety analysis.
     fickling/loader.py:4:from fickling.analysis import
       Severity, check_safety
     fickling/fickle.py:638:module="marshal", attr="loads"
     fickling/analysis.py:10:from fickling.fickle import
       Interpreter, Pickled, Proto
     ...
     (200+ matches across fickling, test, example)
\end{lstlisting}

\vspace{1em}

\begin{lstlisting}[
    basicstyle=\ttfamily\small,
    numbers=none,
    breaklines=true,
    backgroundcolor=\color{orange!15}
]
[reasoning undisclosed]
\end{lstlisting}

\textbf{Tool Call}
\begin{lstlisting}[
    basicstyle=\ttfamily\small,
    numbers=none,
    breaklines=true,
    backgroundcolor=\color{pink!30}
]
CALL: rg -n "pickle" /workspace/repo/fickling \
      /workspace/repo/test /workspace/repo/example | head -200
\end{lstlisting}

\textbf{Observation}
\begin{lstlisting}[
    basicstyle=\ttfamily\small,
    numbers=none,
    breaklines=true,
    backgroundcolor=\color{black!3}
]
  -> example/fault_injection.py:8:import pickle
     example/fault_injection.py:36:Pickled.load(pickle.dumps(...))
     test/test_loads_api.py:2:import pickle
     fickling/hook.py:9:_original_pickle_loads = pickle.loads
     fickling/fickle.py:638:module="marshal", attr="loads"
     ...
     (200 lines, broad matches)
\end{lstlisting}

\vspace{1em}

\begin{lstlisting}[
    basicstyle=\ttfamily\small,
    numbers=none,
    breaklines=true,
    backgroundcolor=\color{orange!15}
]
[reasoning undisclosed]
\end{lstlisting}

\textbf{Tool Call}
\begin{lstlisting}[
    basicstyle=\ttfamily\small,
    numbers=none,
    breaklines=true,
    backgroundcolor=\color{pink!30}
]
CALL: rg -n "loads|load|Unpickler|unsafe|safe|allowed|disallow|blacklist|deny" /workspace/repo/fickling | head -300
\end{lstlisting}

\textbf{Observation}
\begin{lstlisting}[
    basicstyle=\ttfamily\small,
    numbers=none,
    breaklines=true,
    backgroundcolor=\color{black!3}
]
  -> hook.py:16:Drop-in replacement for pickle.Unpickler
       that uses fickling's safety analysis.
     hook.py:27:def load(self):
     hook.py:29:return loader.load(self._file, ...)
     loader.py:8:def load(file, max_acceptable_severity=...):
     loader.py:35:def loads(data, ...):
     analysis.py:332:for node in context.pickled.unsafe_imports()
     fickle.py:39:def is_std_module(module_name): ...
     ...
     (300 lines)
\end{lstlisting}

\vspace{1em}

\begin{lstlisting}[
    basicstyle=\ttfamily\small,
    numbers=none,
    breaklines=true,
    backgroundcolor=\color{orange!15}
]
[reasoning undisclosed]
\end{lstlisting}

\textbf{Tool Call}
\begin{lstlisting}[
    basicstyle=\ttfamily\small,
    numbers=none,
    breaklines=true,
    backgroundcolor=\color{pink!30}
]
CALL: rg -n loads /workspace/repo/fickling | head -100
\end{lstlisting}

\textbf{Observation}
\begin{lstlisting}[
    basicstyle=\ttfamily\small,
    numbers=none,
    breaklines=true,
    backgroundcolor=\color{black!3}
]
  -> hook.py:9:_original_pickle_loads = pickle.loads
     hook.py:55:def new_loads(data, *args, **kwargs):
     __init__.py:2:from .loader import load, loads
     loader.py:24:return pickle.loads(pickled_data.dumps()...)
     loader.py:35:def loads(
     fickle.py:638:module="marshal", attr="loads"
\end{lstlisting}

\vspace{1em}

\begin{lstlisting}[
    basicstyle=\ttfamily\small,
    numbers=none,
    breaklines=true,
    backgroundcolor=\color{orange!15}
]
[reasoning undisclosed]
\end{lstlisting}

\textbf{Tool Call}
\begin{lstlisting}[
    basicstyle=\ttfamily\small,
    numbers=none,
    breaklines=true,
    backgroundcolor=\color{pink!30}
]
CALL: sed -n '1,220p' /workspace/repo/fickling/loader.py;
      echo '---hook';
      sed -n '1,140p' /workspace/repo/fickling/hook.py
\end{lstlisting}

\textbf{Observation}
\begin{lstlisting}[
    basicstyle=\ttfamily\small,
    numbers=none,
    breaklines=true,
    backgroundcolor=\color{black!3}
]
  -> def load(file, max_acceptable_severity=...):
       pickled_data = Pickled.load(file, ...)
       result = check_safety(pickled=pickled_data, ...)
       if result.severity <= max_acceptable_severity:
         return pickle.loads(pickled_data.dumps(), ...)
       raise UnsafeFileError(file, result.to_dict())

     ---hook

     class FicklingSafetyUnpickler:
       def load(self): return loader.load(self._file, ...)

     def run_hook(): pickle.load = loader.load

     def activate_safe_ml_environment(also_allow=None): ...
\end{lstlisting}

\vspace{1em}

\begin{lstlisting}[
    basicstyle=\ttfamily\small,
    numbers=none,
    breaklines=true,
    backgroundcolor=\color{orange!15}
]
[reasoning undisclosed]
\end{lstlisting}

\textbf{Tool Call}
\begin{lstlisting}[
    basicstyle=\ttfamily\small,
    numbers=none,
    breaklines=true,
    backgroundcolor=\color{pink!30}
]
CALL: sed -n '35,110p' /workspace/repo/fickling/loader.py
\end{lstlisting}

\textbf{Observation}
\begin{lstlisting}[
    basicstyle=\ttfamily\small,
    numbers=none,
    breaklines=true,
    backgroundcolor=\color{black!3}
]
  -> def loads(data, max_acceptable_severity=...):
       file = BytesIO(data)
       return load(file, max_acceptable_severity=..., ...)
\end{lstlisting}

\textbf{Tool Call}
\begin{lstlisting}[
    basicstyle=\ttfamily\small,
    numbers=none,
    breaklines=true,
    backgroundcolor=\color{pink!30}
]
CALL: sed -n '1,140p' /workspace/repo/fickling/hook.py
\end{lstlisting}

\textbf{Observation}
\begin{lstlisting}[
    basicstyle=\ttfamily\small,
    numbers=none,
    breaklines=true,
    backgroundcolor=\color{black!3}
]
  -> Full hook.py output, including:
     FicklingSafetyUnpickler
     run_hook
     always_check_safety
     activate_safe_ml_environment
     SafeMLUnpickler
     remove_hook
     deactivate
\end{lstlisting}

\vspace{1em}

\begin{lstlisting}[
    basicstyle=\ttfamily\small,
    numbers=none,
    breaklines=true,
    backgroundcolor=\color{orange!15}
]
[reasoning undisclosed]
\end{lstlisting}

\textbf{Tool Call}
\begin{lstlisting}[
    basicstyle=\ttfamily\small,
    numbers=none,
    breaklines=true,
    backgroundcolor=\color{pink!30}
]
CALL: sed -n '1,260p' /workspace/repo/fickling/analysis.py
\end{lstlisting}

\textbf{Observation}
\begin{lstlisting}[
    basicstyle=\ttfamily\small,
    numbers=none,
    breaklines=true,
    backgroundcolor=\color{black!3}
]
  -> from fickling.fickle import Interpreter, Pickled, Proto

     class AnalyzerMeta(type): ...

     class AnalysisContext:
       def analyze(self, analysis): ...

     class Analyzer(metaclass=AnalyzerMeta): ...

     class Severity(Enum):
       LIKELY_SAFE
       LIKELY_UNSAFE
       LIKELY_OVERTLY_MALICIOUS
       OVERTLY_MALICIOUS

     class AnalysisResult: ...
     class Analysis(ABC): ...
     class DuplicateProtoAnalysis(Analysis): ...
     class MisplacedProtoAnalysis(Analysis): ...

     (first 260 lines; does not reach OvertlyBadEvals)
\end{lstlisting}

\vspace{1em}

\begin{lstlisting}[
    basicstyle=\ttfamily\small,
    numbers=none,
    breaklines=true,
    backgroundcolor=\color{orange!15}
]
[reasoning undisclosed]
\end{lstlisting}

\textbf{Tool Call}
\begin{lstlisting}[
    basicstyle=\ttfamily\small,
    numbers=none,
    breaklines=true,
    backgroundcolor=\color{pink!30}
]
CALL: rg -n "unsafe|Danger|Bad|safety|Severity|GLOBAL|import|allowed|SAFE|OVERT|LIKELY" /workspace/repo/fickling/analysis.py | head -200
\end{lstlisting}

\textbf{Observation}
\begin{lstlisting}[
    basicstyle=\ttfamily\small,
    numbers=none,
    breaklines=true,
    backgroundcolor=\color{black!3}
]
  -> 69:class Severity(Enum):
     117:Returns True if there is no evidence ... unsafe
     186:class InvalidOpcode(Analysis):
     203:"inherently unsafe"
     249:class UnsafeImportsML(Analysis):
     283:class BadCalls(Analysis):
     296:class OvertlyBadEvals(Analysis):
     324:"can execute arbitrary code and is inherently unsafe"
     332:for node in context.pickled.unsafe_imports()
     ...
     (58 matches)
\end{lstlisting}

\vspace{1em}

\begin{lstlisting}[
    basicstyle=\ttfamily\small,
    numbers=none,
    breaklines=true,
    backgroundcolor=\color{orange!15}
]
[reasoning undisclosed]
\end{lstlisting}

\textbf{Tool Call}
\begin{lstlisting}[
    basicstyle=\ttfamily\small,
    numbers=none,
    breaklines=true,
    backgroundcolor=\color{pink!30}
]
CALL: rg -n "unsafe" /workspace/repo/fickling/analysis.py
\end{lstlisting}

\textbf{Observation}
\begin{lstlisting}[
    basicstyle=\ttfamily\small,
    numbers=none,
    breaklines=true,
    backgroundcolor=\color{black!3}
]
  -> 117:"Returns True if there is no evidence ... unsafe"
     203:"inherently unsafe"
     240:"torch.load() which is unsafe..."
     324:"can execute arbitrary code and is inherently unsafe"
     332:for node in context.pickled.unsafe_imports()
\end{lstlisting}

\vspace{1em}

\begin{lstlisting}[
    basicstyle=\ttfamily\small,
    numbers=none,
    breaklines=true,
    backgroundcolor=\color{orange!15}
]
[reasoning undisclosed]
\end{lstlisting}

\textbf{Tool Call}
\begin{lstlisting}[
    basicstyle=\ttfamily\small,
    numbers=none,
    breaklines=true,
    backgroundcolor=\color{pink!30}
]
CALL: sed -n '180,350p' /workspace/repo/fickling/analysis.py
\end{lstlisting}

\textbf{Observation}
\begin{lstlisting}[
    basicstyle=\ttfamily\small,
    numbers=none,
    breaklines=true,
    backgroundcolor=\color{black!3}
]
  -> ERROR: Terminal call budget exhausted (15/15).
     Please submit your answer.
\end{lstlisting}

\vspace{1em}

\textbf{Retry Messages}
\begin{lstlisting}[
    basicstyle=\ttfamily\small,
    numbers=none,
    breaklines=true,
    backgroundcolor=\color{black!3}
]
Please use one of the available tools:
terminal, submit_vulnerable_files, or
submit_no_vulnerability_found.

Repeated for 3 retries.
\end{lstlisting}

\vspace{1em}

\begin{lstlisting}[
    basicstyle=\ttfamily\small,
    numbers=none,
    breaklines=true,
    backgroundcolor=\color{orange!15}
]
[reasoning undisclosed]
\end{lstlisting}

\textbf{Final Tool Call}
\begin{lstlisting}[
    basicstyle=\ttfamily\small,
    numbers=none,
    breaklines=true,
    backgroundcolor=\color{pink!30}
]
CALL: submit_vulnerable_files
\end{lstlisting}

\textbf{Submission}
\begin{lstlisting}[
    basicstyle=\ttfamily\small,
    numbers=none,
    breaklines=true,
    backgroundcolor=\color{blue!8}
]
  -> ["fickling/analysis.py", "fickling/loader.py"]
\end{lstlisting}

\vspace{1em}

\textbf{Result}
\begin{lstlisting}[
    basicstyle=\ttfamily\small,
    numbers=none,
    breaklines=true,
    backgroundcolor=\color{mushygreen!20}
]
Precision: 0.5
Recall:    0.5
File F1:   0.5
\end{lstlisting}

\end{tcolorbox}